\documentclass[pre,preprint,preprintnumbers,amsmath,amssymb]{revtex4}
\usepackage{graphics,epstopdf,epsfig,stmaryrd}

\newcommand{\mX}{\mathcal{X}}

\newcommand{\mx}{\mathbf{x}}
\newcommand{\me}{\mathbf{E}}
\newcommand{\mE}{\mathbf{E}_1}
\newcommand{\mk}{\mathbf{k}}
\newcommand{\res}{Eq.~(\ref{res})}
\newcommand{\si}{\langle \sin(\psi) \rangle}

\newcommand{\mL}{\mathcal{L}}
\newcommand{\mF}{\mathcal{F}}
\newcommand{\mH}{\mathcal{H}}
\newcommand{\mh}{\mathcal{E}}
\newcommand{\resun}{Eq.~(\ref{res1})}
\newcommand{\reso}{Eq.~(\ref{res0})}
\newcommand{\lL}{\langle L \rangle}
\newcommand{\mI}{\mathcal{I}}
\newcommand{\mA}{\mathcal{A}}

\begin {document}
\title{Envelope equation for the linear and nonlinear propagation of an electron plasma wave, including the effects of Landau damping, trapping, plasma inhomogeneity,  and the change in the state of wave}
\author{Didier B\'enisti}
\email{didier.benisti@cea.fr}
\affiliation{ CEA, DAM, DIF F-91297 Arpajon, France.}

\date{\today}
\begin{abstract}
This paper addresses the linear and nonlinear three-dimensional propagation of an electron wave in  a collisionless plasma that may be inhomogeneous, nonstationary, anisotropic and even weakly magnetized. The wave amplitude, together with any hydrodynamic quantity characterizing the plasma (density, temperature,Ê\dots) are supposed to vary very little within one wavelength or one wave period. Hence, the geometrical optics limit is assumed, and the wave propagation is described by a first order differential equation. This equation explicitly accounts for  three-dimensional effects, plasma inhomogeneity, Landau damping, and the collisionless dissipation and electron acceleration due to trapping. It is derived by mixing results obtained from a direct resolution of the Vlasov-Poisson system and from a variational formalism involving a nonlocal Lagrangian density. In a one-dimensional situation, abrupt transitions are predicted in the coefficients of the wave equation. They occur when the state of the electron plasma wave changes, from a linear wave to a wave with trapped electrons. In a three dimensional geometry, the transitions are smoother, especially as regards the nonlinear Landau damping rate, for which a very simple effective and accurate analytic expression is provided.
\end{abstract}
\maketitle
\section{Introduction}
\label{I}
Electron oscillations are one of the first  effects ever reported in plasma physics, by Tonks and Langmuir~\cite{langmuir}, nearly one century ago. Since then, a countless number of publications have shown than the propagation and dispersion relation of these oscillations, henceforth named electron plasma waves (EPW's), were usually more complex than those of the so-called Langmuir waves described in Ref.~\cite{langmuir}. In particular, kinetic effects, involving wave-particle interaction at a microscopic level, are often to be accounted for. For example, in the linear regime, these are well-known to be responsible for Landau damping~\cite{landau,comment,mouhot}.  Moreover, in a realistic physics situation, the plasma is never homogeneous nor stationary. Hence, microscopic effects have to be solved together with the long time and length scale evolution of the plasma. This makes the study of EPW's particularly difficult, even when addressed within some limit. 

Very often, the geometrical optics limit happens to be relevant.  It applies to waves whose electric field, $\me \equiv \me[\mathbf{x},t,\psi(\mathbf{x},t)]$, is $2\pi$-periodic with respect to the eikonal, $\psi$ \cite{comment2} (when varied at fixed $\mathbf{x}$ and $t$), and varies slowly with $\mathbf{x}$ and $t$ (at fixed $\psi$). Moreover, the wave frequency, $\omega \equiv -\partial_t \psi$, and wave number, $\mk\equiv\mathbf{\nabla}Ê\psi$, together with the  wave phase velocity, $v_\phi \equiv \omega/\vert \mathbf{k} \vert$, are also assumed to vary much more slowly than $\psi$, which may only be true if the hydrodynamic properties of the plasma (its density, temperature, \dots) are slowly-varying too. When the wave electric field is not sinusoidal, all its Fourier components in $\psi$ may be expressed in terms of the first one, $\mE$, as described in Ref.~\cite{lindberg}. Then, within the geometrical optics limit, the dispersion relation is written as an equation between $\omega$, $k\equiv\vert \mk \vert$ and $E_1\equiv\vert \mE \vert$, which needs to be solved together with the consistency relation, $\partial_t \mk=-\mathbf{\nabla}Ê\omega$, and appropriate boundary conditions. As for the evolution of $\mE$, it is derived from a first order partial differential equation, henceforth named envelope equation. Now, in spite of the very numerous publications on EPW's, no derivation of an envelope equation, valid in the linear and nonlinear regimes, and in a three-dimensional (3-D) inhomogeneous  plasma, could be found in previous publications. The purpose of this paper is, precisely, to fill this gap, by providing the wave equation~(\ref{res}), describing the propagation of EPW's in a collisionless plasma whose unperturbed distribution function varies slowly  in space, time, and velocity. Moreover, in addition to assuming slow density variations, we also need to exclude the situation when these slow variations are either random or periodic, so that Anderson localization~\cite{escande,doveil} is not expected to occur. In particular, the density is not modulated by any ion wave, and the EPW is supposed to be the only electrostatic mode propagating in the plasma. The situation when two counterpropagating lasers would produce a transient photonic crystal~\cite{lehmann} is, therefore, also excluded. Note, however, that Anderson localization has only been derived in the linear regime while, as discussed in the conclusion, well-known linear limitations of a first order envelope equation, such as diffraction or group velocity dispersion, may become irrelevant due to nonlinear effects. Hence, the range of validity of~\res~is usually different from what could be inferred from known linear results. This point is further detailed in the conclusion, where we mainly focus of nonlinear effects that have been observed both numerically and experimentally to discuss the range of validity of our main result,~\res.

Now, instead of expressing all Fourier components of the electric field as a function of the first one in order to derive an envelope equation for $\mE$, it is more natural, and more simple, to write this equation in terms of one single nonlinear function. This is what we do in this paper when resorting to a variational formalism. Moreover, we discuss the ability to write the EPW envelope equation in terms of what could be defined as the wave action. A definition for the wave action is proposed in Paragraph~\ref{new4}, and its variations are investigated in connection with the concept of plasmon.

Our main result, Eq.~(\ref{res}), models the propagation of an electron plasma wave that has grown from the noise level, until it has (possibly) entered the nonlinear regime. Moreover, we restrict to the situation when the EPW is localized in one given space region, which allows to address the electron response to the EPW in quite a simple fashion. Indeed, we show that this response mainly depends of the duration, $t_{int}$, of the interaction with the wave, compared to the bounce period, $T_B$, which is the period of a trapped orbit at the bottom of the potential. If $t_{int}Ê\alt T_B$, the electron motion is little affected by the EPW, and is well described by a perturbation analysis in the wave amplitude. As discussed in Section~\ref{new}, first order results are already quite accurate,  so that the electron response to the wave remains close to linear when $t_{int}Ê\alt T_B$. 
When $t_{int}Ê\agt T_B$, and for the slowly-varying wave considered in this article, the electron distribution function has been efficiently phase mixed, so that the electron response may be considered as nearly adiabatic. By nearly adiabatic, we mean here that the distribution in angle, canonically conjugated to the dynamical action defined in Appendix~\ref{A}, may be considered as uniform over one wavelength. Moreover, the slowness required to reach the near-adiabatic regime has to be related to the smoothness of the distribution function. This reads $\vert \gamma \vert/k \ll \Delta v$, where $\gamma$ is the typical rate of variation of the EPW, as experienced by the electrons, and $\Delta v$ is the typical velocity range of variation of the unperturbed distribution function (hence, for an initially Maxwellian plasma, the condition just reads $\vert \gamma \vert/kv_{th}Ê\ll 1$, where $v_{th}$ is the thermal velocity). Furthermore, as the wave grows, trapping has to be effective to reach the near-adiabatic regime.  As discussed in Section \ref{new}, this requires that, at least by the time phase mixing has occurred, $\vert dv_\phi/dt \vert < (4/\pi) \vert d(\omega_B/k)/dt \vert$, where $\Omega_B\equiv 2\pi/T_B$ and where the derivatives are calculated along the electrons motion in the wave frame. Now, as shown in Section~\ref{new}, the transition between the linear and near-adiabatic responses occurs very abruptly when $t_{int}Ê\approx T_B$, so that one may consider that there only exist two classes of electrons in the plasma, linear and near-adiabatic ones. This result is extremely important because it solves, in a very simple fashion, the issue of wave-particle interaction. In particular, it allows to provide a very simple and quite accurate expression for the nonlinear collisionless damping rate, $\nu_{NL}$, which reads $\nu_{NL}=\eta_{lin}\nu_L $, where $\nu_L$ is the linear Landau damping rate, and $\eta_{lin}$ is the fraction of linear electrons. Moreover, we are able to provide an upper and lower bound for $\eta_{lin}$ which are very close to each other. Hence, although addressing nonlinear collisionless damping in a complete fashion is quite  a difficult issue (as discussed in Paragraph~\ref{new2}), we nevertheless derive in this paper a very simple and precise estimate for $\nu_{NL}$.

Now, the latter results on the microscopic electron motion need to be coupled with the variations of the wave and of the plasma over long space and time scales. Variational methods are particularly suited to do so, and have become very popular to study wave propagation in an inhomogeneous medium after Whitham's work on the subject~\cite{whitham}. They are indeed very powerful, because it is enough to calculate the Lagrangian density at zero order in the variations of the medium to derive, automatically, an envelope equation that accounts for these variations at first order. Recently, a variational formalism has been introduced in Refs.~\cite{dodin12,dodin13b}~to study the EPW propagation in an inhomogeneous and nonstationary plasma, within two hypotheses ; (i) all electrons are assumed to be adiabatic ; (ii) their orbits are assumed to never come close to the frozen separatrix. In this paper, we work again on the variational formalism introduced in Refs.~\cite{dodin12,dodin13b}, and we extend its range of validity by allowing for separatrix crossing. Moreover, we use the results of Ref.~\cite{benisti15} to also account for the response of linear electrons in the envelope equation. In a one dimensional (1-D) geometry, this yields Eq.~(\ref{63}), describing the abrupt transition from a linear to near-adiabatic wave propagation. In particular, the way collisionless dissipation is changed from Landau damping to the dissipation induced by trapping is discussed in detail in Paragraph~\ref{new1}. Moreover, remarkably enough, we can identify a term in the envelope equation that is zero in the linear regime and abruptly reaches a finite value once electron trapping has become effective. This term accounts for the fact that, when some electrons have been trapped in the wave potential, a change in phase velocity accelerates (or decelerates) them at the expense (or benefit) of the wave. Clearly, this effect has no linear counterpart. Similarly, as already discussed in Ref.~\cite{dodin12}, when the plasma density is not uniform, the inhomogeneous loading of trapped electrons yields a term in the envelope equation which has no linear counterpart either. These changes in the envelope equation bear some analogy with phase transition when one identifies an order parameter that is identically zero in one phase and assumes a finite value once the phase has changed. Hence, there is a transition in the way the EPW propagates, as though the state of this wave was changing, from a linear wave to a wave with trapped electrons. In 1-D, this change of state is very abrupt, as described by Eq.~(\ref{63}). In 3-D, the transition is much smoother because, at any time and space location, there always is a mixture of linear and near-adiabatic electrons. Consequently, our 3-D envelope equation,~\res, simply reads as a weighted sum of a linear and of a near-adiabatic wave equation. It accounts for a very rich physics, that is discussed at the end of Paragraph~\ref{IV.2}. In particular, as already mentioned above, collisionless damping is allowed for through an effective damping rate, for which we provide a particularly simple and accurate analytical expression. Actually, we believe that~\res~accurately accounts for all effects entering into the EPW propagation, when the unperturbed distribution function is smooth enough, and within the limits of geometric optics which we further discuss at the end of our paper. 

This article is organized as follows. In Section~\ref{III}, we address the plasma wave propagation within two different limits, the linear and near-adiabatic one. In Paragraph~\ref{III.1},  we recall the results on wave propagation derived in Ref.~\cite{benisti15}~when all electrons are linear. In Paragraph~\ref{III.2.1}, the results of Refs.~\cite{dodin12,dodin13b} when all electrons are nearly adiabatic are rederived in a slightly different way. Then, the range of validity of these results is extended in order to allow for orbits arbitrarily close to the separatrix.  Moreover, in Paragraph~\ref{III.2.2}, we show how to account for separatrix crossing by making use of a non local variational formalism. In Section~\ref{new}, we discuss how the linear (or perturbative) and the adiabatic results may be connected. In Paragraph~\ref{new1}, we investigate how the wave equation changes during the transition from a linear to a near-adiabatic propagation. In Paragraph~\ref{new2}, we indicate how to actually make  the connection, and in Paragraph~\ref{new3}, we test our theory against numerical simulations. From these results, we derive the EPW envelope equation, first in 1-D, in Paragraph~\ref{IV.1},  and then in 3-D, in Paragraph~\ref{IV.2}. This yields our main result, \res. The ability to write this equation in terms of the wave action is, moreover, discussed in Paragraph~\ref{new4}. Finally, Section~\ref{V} concludes this work, discusses its limits together with the way it could be applied or tested experimentally. Each Section of this manuscript starts with a brief summary of its main results, so that one may understand each step leading to~\res, without going through all the details of the reasoning.

\section{Linear  and near-adiabatic envelope equations}
\label{III}
In this Section, we provide two envelope equations obtained in two different limits, the linear and near-adiabatic ones. The linear envelope equation is \resun. It has been derived in Ref.~\cite{benisti15}, from a direct resolution of the Vlasov-Gauss system, although, as shown in Appendix~\ref{B}, Landau damping may also be recovered from a variational principle. As regards the EPW propagation in the adiabatic regime, it is ruled by Eq.~(\ref{51}), derived by making use a variational formalism similar to that introduced by Dodin and Fisch in Ref.~\cite{dodin12}. Moreover, the results of Ref.~\cite{dodin12} have been generalized in order for Eq.~(\ref{51}) to allow for separatrix crossing.  As discussed in Section~\ref{new}, Eqs.~(\ref{res1})~and~(\ref{51}) are enough  to derive an accurate description of wave propagation in a general situation. 

\subsection{Linear limit}
\label{III.1}
In this Paragraph, we recall the envelope equation derived in Ref.~\cite{benisti15} in the linear limit. The plasma is characterized by its unperturbed distribution function, $f_H(\mathbf{x},\mathbf{v},t)$, such that $\int f_H d\mathbf{v} =n$, where $n$ is the local electron density. $f_H$ obeys the following Boltzman equation,
\begin{equation}
\label{18}
\frac{\partial f_H}{\partial t}+\mathbf{v}.\frac{\partial f_H}{\partial \mathbf{x}}+ \frac{\mathbf{F_H}}{m}.\frac{\partial f_H}{\partial \mathbf{v}}=C\{f_H\},
\end{equation}
where $\mathbf{F_H}$ is a force field (not including the effect of the wave) that makes the plasma slowly evolve in space and time, and where $C$ is a collisional operator. Then, the EPW envelope equation is found to be
\begin{equation}
\label{res1}
\partial_t \partial_\omega \Lambda_{lin}-\partial_{\mx}.\partial_{\mk}Ê\Lambda_{lin}+2(\nu_L+\nu_c)\partial_\omega \Lambda_{lin}=\varepsilon_0 E_1E_d/2,
\end{equation}
where $E_d$ is some drive amplitude (that may be zero), and where $\Lambda_{lin}Ê\equiv \varepsilon_0\partial_\omega \chi_{lin}E_1^2/4$, with,
\begin{equation}
\label{20}
\chi_{lin} = -\frac{\omega_{pe}^2}{nk} P.P.\left( \int \frac{f^{'}_H}{kv_k-\omega}dv_k\right).
\end{equation}
In Eq.~(\ref{20}), $v_k \equiv \mathbf{k}.\mathbf{v}/k$ and $f'_H \equiv \partial_{v_k}f_H$. As for the collisional damping rate, $\nu_c$, it is,
\begin{equation}
\label{22}
\nu_c = \frac{\omega_{pe}^2}{nk} P.P. \left[\int C\left\{\frac{f'_H}{kv_k-\omega}\right\}\frac{dv}{kv_k-\omega}\right],
\end{equation}
while the Landau damping rate, $\nu_L$, is found to be,
\begin{equation}
\label{21}
\nu_L \equiv -\frac{\pi \omega_{pe}^2}{ nk^2Ê\partial_\omega \chi_{lin} }f'_H(x,v_\phi,t).
\end{equation}
Hence, there is no first order correction to $\nu_L$ due to the space or time variations of plasma density, nor due to the variations of the wave amplitude, wave number and wave frequency.

 Eq.~(\ref{res1}) has been derived from a direct resolution of the Vlasov-Gauss system while, in Paragraph~\ref{III.2}, we introduce a variational principle to derive the envelope equation in the near-adiabatic limit. We would like to stress here that Landau damping may also be derived from a variational principle, using a non local Lagrangian density. We show this in Appendix~\ref{B}, in the very simple situation when the plasma and the wave amplitude are space-independent, and when $k$ and $\omega$  are constant. A more general result is certainly achievable, but our point is not provide a lengthy derivation of an already published result, but, simply, to show the ability of variational methods to allow for damping.

Moreover, when the wave is homogeneous, when collisions may be neglected and when the wave is not driven, \resun~reads,
\begin{equation}
\label{a1_0}
d_t\partial_\omega \Lambda_{lin}+2\nu_L \partial_\omega\Lambda_{lin}=0,
\end{equation}
 and this equation is often interpreted as the Landau damping of the linear wave action density, $\mI_{lin} \equiv  \partial_\omega\Lambda_{lin}$. If $\mI_{lin}$ is associated to a plasmon density, Eq.~(\ref{a1_0})~may be seen as plasmon conversion into electron kinetic energy. Now, from the results of Appendix~\ref{B}, Eq.~(\ref{a1_0})~is equivalent to,
\begin{equation}
\label{a2}
d_t\partial_\omega \mL_{lin}=0,
\end{equation}
where $\mL_{lin}$ is defined by Eq.~(\ref{B15}). This equation shows that, unlike what is usually assumed in variational formalisms, the frequency derivative of the Lagrangian density may not always be defined as the wave action density. Indeed, from Eq.~(\ref{a2}), $\partial_\omega \mL_{lin}$ remains constant, and it would not make quite sense to define a constant action for a wave which is Landau damped and, therefore, whose amplitude keeps decreasing. 

\subsection{Near-adiabatic limit}
\label{III.2}

The adiabatic limit usually refers to the situation when the electron dynamical action, $I$, precisely defined in Appendix~\ref{A}, may be considered as a constant. In the homogeneous situation when the wave amplitude only depends on time, it has been proved in Ref.~\cite{tim,cary,hanna,nei,ten,vas},  that, indeed, $I$ changes very little when the dynamics is slowly-varing, even after the electrons have crossed the frozen separatrix. Then, making the so-called adiabatic approximation amounts to neglecting the change in $I$ and to assuming that the distribution in the angle, $\theta$, canonically conjugated to $I$, is uniform. This assumption is discussed in great detail in Section~\ref{new}, where it is shown to be valid only once $\int \omega_B dt$ (where $\omega_B \equiv 2\pi/T_B$), is large enough, and it is related to phase mixing. Moreover, the numerical results of Paragraph~\ref{new3} indicate that $\int \omega_B dt \agt 5$ is enough. This value is close to $2\pi$, which would correspond to the time it takes to a trapped electron to complete its orbit, which makes sense. Indeed, the adiabatic approach explicitly accounts for a population of trapped electrons and, therefore, it may only be valid once trapping has become effective.

When the wave amplitude is space-dependent, $I$ is not an adiabatic invariant, it slowly varies with time, as derived in Appendix~\ref{A}. Consequently, $\theta$ is not uniformly distributed. However, in what we call ``the near-adiabatic limit'', the  electron distribution in $\theta$ may be considered as uniform \textit{within one wavelength}. Again, this limit is reached when $\int \omega_B dt \agt 5$, the integral being calculated along any electron orbit, in the wave frame. Clearly, only in 1-D may $\int \omega_B dt$ be larger than $5$ for all electrons. Hence, in this Section, we restrict to the 1-D limit while the generalization to a three-dimensional geometry will be made in Section~\ref{IV}.

\subsubsection{Envelope equation ignoring the effect of separatrix crossing}
\label{III.2.1}

In this Paragraph, we detail the averaged variational formalism valid in the near-adiabatic regime. The derivation is made in the same spirit as in Ref.~\cite{dodin13b}, leading to the same results. However, we use here a space averaging, instead of a temporal one, which allows to extend the range of validity of the wave equation~(\ref{res0}). Namely, unlike in Ref.~\cite{dodin13b}, we show that this equation is valid even when some electron orbits are located close to the frozen separatrix. 

We focus here on the evolution of the plasma wave. Hence, in the Lagrangian density ruling the evolution of the electrons, and of the wave, we only retain the electric energy density due to the EPW. However, the Lagrangian for the electron motion allows for any external field that may not be due to the wave. In other words, only the electrons and the EPW are treated kinetically. All other fields, as well as the ion response, are assumed to derive from macroscopic, fluid-like, equations. Actually, one could easily include any field or type of particle in the Lagrangian density. However, this would lead to unnecessary complications in the situation considered in this paper where only one electron mode exists in addition to slowly varying, fluid-like, fields. Moreover, we adopt a Vlasovian approach and introduce the electron distribution function, $f(x,v,t)$. Then, the Lagrangian density, expressed as a function of $x$ and $t$, is, 
\begin{equation}
\label{23}
L(x,t)=\frac{\varepsilon_0 E^2}{2}+\int L_e (x,v,t) f(x,v,t)dv,
\end{equation}
where
\begin{equation}
\label{23b}
 L_e (x,v,t) \equiv  \frac{mv^2}{2}-evA(x,t)-\Phi(x,t).
\end{equation}
In Eq~(\ref{23}), the wave electric field is $E \equiv -\partial_t (A-A_{ext})-\partial_x (\Phi-\Phi_{ext})$. Hence, in the electron Lagrangian, $L_e$, the external potentials, $\Phi_{ext}$ and $A_{ext}$, account for any slowly varying electric field, or weak magnetic field, that may exist in the plasma and that is not due to the wave. Moreover, as noted in Ref.~\cite{lindberg}, it is important to allow for a homogeneous vector potential, $\delta A \equiv (A-A_{ext})$, in the wave electric field in order to correctly calculate the dispersion relation of large amplitude EPW's. As for the wave potential, it reads $\Phi-\Phi_{ext}Ê\equiv \Phi_0+\sum_n \Phi_n\sin(n\psi+\delta \phi_n)$, where each $\Phi_n$ and $\delta \phi_n$ varies slowly in space and time. 

Now, in the spirit of the multiple-scale analysis, $L$ reads, $L \equiv L(\psi,x,t)$ and is periodic in $\psi$ (when it is varied at constant $x$ and $t$). Then, following Whitham~\cite{whitham}, we introduce the $\psi$-averaged value of $L$, which we denote by $\langle L \rangle$. It is just the value of $L_e$, averaged over all electrons located within one wavelength about a given position, to which is added the space averaged value of $\varepsilon_0 E_1^2/2$. From Whitham, one knows that $\lL$ may be used as a regular Lagrangian density to describe the average evolution of the wave and particles. In particular, from Ref.~\cite{whitham}, one would expect $\partial_{xk} \lL -\partial_{t\omega} \lL=0$. However, this equation is not correct here. Indeed, as discussed in Refs.~\cite{dodin12,seliger}, when using an Eulerian description for the distribution function, as in Eq.~(\ref{23}), one needs to introduce Lagrange multipliers that account for the constraint imposed by the conservation of the distribution function. One way to alleviate this difficulty is to write the distribution function in terms of Lagrangian variables, i.e., to consider $f$ as a function the electron positions and velocities (see Appendix \ref{C} for more details). Now, as discussed in Appendix~\ref{A}, the coordinates of the trapped electrons are necessarily defined with respect to the position of the $O$-point of their trapping island. This makes $f$, or, more precisely, the trapped electrons distribution function, $f_t$, an explicit function of $\psi$ [in particular, see Eq.~(\ref{A6b}) and the discussion below Eq.~(\ref{31})]. This needs to be accounted for when writing Lagrange equations for the EPW electric field. Moreover, due to the eikonal dependence of $f_t$, the Lagrangian density is no longer periodic in $\psi$ and Whitham theory no longer directly applies. Nevertheless, we show in Appendix~\ref{C} that, $\partial_{xk} \lL -\partial_{t\omega} \lL = \partial_\psi \lL$, so that $\lL$ may indeed be viewed as a regular Lagrangian density. Actually, as discussed in Appendix~\ref{C}, the former Lagrange equation is only approximately true, valid only at lowest order in the variations of $f_t$. 

 In order to derive $\lL$, we proceed as in the Appendix~\ref{C} and shift to action-angle variables. Hence, let us denote by $\tilde{f}_u$ the distribution function of the untrapped electrons in $I$ and $\theta$ (the distribution function in action-angle variables, $\tilde{f}$, is defined such that $\tilde{f}dId \theta=fdxdv$, so that $\tilde{f}=f/m$). As long as $I$ remains nearly conserved, $\tilde{f}_u$ varies very slowly with $\theta$. Actually, in the near-adiabatic regime, the $\theta$-dependence of $\tilde{f}_u$ is only due to the small plasma inhomogeneity. Therefore, $\tilde{f}_u $ reads, $\tilde{f}_u \equiv \tilde{f}_u[x(\theta,I),I,t]$, and  varies very slowly with $x$. At the order where we work, $\tilde{f}_u$ may be considered as uniform within one wavelength. Then, if one denotes by $\tilde{L}_e$ the Lagrangian for wave-particle interaction expressed in action-angle variables,  the untrapped electrons provide a contribution to $L$ which we denote by $L_u$, and which is, 
 \begin{equation}
\label{n15}
 L_u(x_0,t)= \int \tilde{L}_e \partial_x \theta\vert_{\theta_0} \tilde{f}_udI,
\end{equation}
where $\theta_0$ is such that $x(\theta_0,I)=x_0$. Now, using the results of Appendices~\ref{A}~and~\ref{C}, 
\begin{equation}
\label{n16}
\tilde{L}_e \approx I\Omega-\mh+mu^2/2+e^2A^2/2m,
\end{equation}
where,
\begin{equation}
\label{n17}
\mh \equiv m(v-u)^2/2+\Phi,
\end{equation}
is related to $I$ by Eq.~(\ref{A9}), where $\Omega \equiv \partial_I \mh$, and where $u\equiv v_\phi-eA/m$. Moreover, it is clear that, in order to derive the Lagrange equations for the fields, those entering in $\tilde{L}_e$ have to be expressed in terms of $x$ and $t$. Then, considering $\tilde{L}_e$ as a slowly-varying function of $x$ and $t$, and  using $\partial_x \theta=k \partial_\psi \theta$, one finds,
\begin{equation}
\label{24}
L_u(x_0,t)= \int k \tilde{L}_e(x_0,t) \tilde{f}_u \partial_\psi \theta dI.
\end{equation}
Following Whitham, we now average $L_u$ over $\psi$ while keeping $x_0$  constant. In the integral of~Eq.~(\ref{24}), this amounts to averaging over $\theta$ while keeping all factors, except $\partial_\psi \theta$, constant. Then, since $\langle \partial_\psi \theta \rangle=1$, one straightforwardly finds, 
\begin{equation}
\label{25}
\langle L_u \rangle (x_0,t)= \int k \tilde{L}_e(x_0,t)\tilde{f}_u  dI.
\end{equation}
One recovers the value found for $\langle L_u \rangle$ in Refs.~\cite{dodin12,dodin13b}, except that we now refer to a space averaging (instead of a time averaging). Consequently, as discussed in Appendix~\ref{A}, our expression for $\langle L_u \rangle$ is valid even when some electron orbits are arbitrarily close to the frozen separatrix. 

 Now, remember that $\tilde{f}_u \equiv \tilde{f}_u(x,I,t)$ and, at the order where we work, one may choose for $x$ any position located within one wavelength from $x_0$. Henceforth, we choose to evaluate $\tilde{f}_u$ at $X= \langle x(\theta,I) \rangle_\theta$, where $\langle . \rangle_\theta$ denotes a $2\pi$-averaging in $\theta$ which is, here, performed about $\theta_0$ such that $x(\theta_0,I)=x_0$. $X$ is the so-called oscillation center~\cite{dewar73}. Moreover, following Ref.~\cite{dodin12}, we introduce the oscillation center velocity, $V \equiv dX/dt = \langle \dot{x}(\theta,I) \rangle_\theta$, and $P \equiv \partial  \tilde{L}_e /\partial V$. Then, as shown in Ref.~\cite{dodin12}, $P=kI$ and $\tilde{L}_e= PV-\mH_u$, where 
\begin{equation}
\label{27}
\mH_u=\mh+Pv_\phi-mv_\phi^2/2-eAv_\phi.
\end{equation}
Since we showed in Appendix~\ref{A}  that $\langle dP/dt \rangle=-\partial_X\mH_u$, we conclude that, provided that $P$ varies very little within one wavelength, $X$ and $P$ are canonically conjugated variables for $\mH_u$. Hence, denoting $f_u(X,P,t) \equiv \tilde{f}_u[X,I(P),t]$ one finds,
\begin{equation}
\label{28}
\langle L_u \rangle = \int (PV-\mH_u) f_u(X,P,t) dP.
\end{equation}
Clearly, the term $PV$ in Eq.~(\ref{28}) does not enter into the Lagrange equations for the wave. Hence, as regards the wave evolution, one just needs to consider
\begin{equation}
\label{29}
\mL_u   \equiv -\int \mH_u f_u(X,P,t) dP.
\end{equation}
As for $f_u$, since $X$ and $P$ are canonically conjugated for $\mH_u$, one finds 
\begin{equation}
\label{30}
\partial_t f_u+\partial_P \mH_u \partial_X f_u -\partial_X \mH_u \partial_P f_u=0.
\end{equation}
If one were to use Eq.~(\ref{30}) over long time scales, one would need to account for collisions. Deriving the collision operator in $(X,P)$ coordinates  is, however, out of the scope of this paper. Moreover, our theory is only valid provided that $f_u$ remains slowly varying in $X$, and Eq.~(\ref{30}) is mainly useful to tell whether this property remains, indeed, fulfilled. Note also that a geometric change in $I$ as that described in Ref.~\cite{ratchet}, or due to symmetric detrapping, as explained in Ref.~\cite{dissipation}~and Paragraph~\ref{III.2.2}, entails a discontinuity in the $I$-variations of $f_u$ which makes the use of Eq.~(\ref{30}) questionable. However, as explained in Appendix~\ref{A}, this equation needs to be understood as an equation for the averaged value of $f_u$, over a small interval in $I$ of the order of $m \vert \gamma \vert/k^2$, where $\gamma$ is the typical growth rate of the EPW. This smoothes out the discontinuity.

As regards the trapped electrons, $\tilde{L}_e$ is the same as for the untrapped ones, and reads $\tilde{L}_e=I\Omega-\mH_t$, where
\begin{equation}
\label{31}
\mH_t=\mh-mu^2/2-e^2A^2/2m.
\end{equation}
Now, as discussed in Appendix~\ref{A}, for trapped electrons, there is one set of action-angle variables for each resonance (i.e., each trapping island). These are labelled by, $\psi_O$, the value of the wave eikonal at the $O$-point of the resonance. Then, the trapped distribution function reads, $\tilde{f}_t \equiv \tilde{f}_t(\Psi_O,I,t)$. It is uniform within one wavelength (for each $I$), and it is conserved in the wave frame, $\tilde{f}_t[\psi_O(t),I]=\tilde{f}_t[\psi_O(0),I]$ (since, as shown in Appendix~\ref{A}, the action of trapped electrons is an adiabatic invariant). Hence, following the same steps as for the untrapped electrons, one finds that the contribution to $\lL$ from the trapped electrons, which we denote by $\langle L_t \rangle$, is
\begin{equation}
\label{32}
\langle L_t \rangle  \equiv \int (I\Omega-\mH_t) \tilde{f}_t(\psi,I) d(kI),
\end{equation}
where $\tilde{f}_t$ appears as an explicit function of $\psi$. Therefore, unlike for the untrapped electrons, when applying the least action principle, one needs to account for the variations of $\tilde{f}_t$ with respect to $\psi$. Moreover, since the term $I\Omega$ in Eq.~(\ref{32}) does not enter into the wave evolution, one may just consider
\begin{equation}
\label{33}
\mL_t \equiv -\int \mH_t k\tilde{f}_t(\psi,I) dI.
\end{equation}
Therefore, following~Ref.~\cite{dodin12}, we introduce the Routhian density,
\begin{equation}
\label{34}
\mL \equiv \varepsilon_0\langle E^2 \rangle/2+\mL_u+\mL_t,
\end{equation}
and, from Appendix~\ref{C}, we know the wave equations may be derived from a blind use of Lagrange equations on $\mL$. When these are applied to the field, $\psi$, they yield, $\partial^2_{t\omega}Ê\mL-\partial^2_{xk}Ê\mL=\int \mH_t \partial_\psi \tilde{f}_t kdI$. In the remainder of this paper, we prefer expressing the trapped distribution function in terms of $x$ and $I$, and we introduce, $f_t(x,I,t) \equiv \tilde{f}_t[\psi(x,t),I]$. Then, using $\partial_\psi \tilde{f}_t=k^{-1}Ê\partial_x f_t$, one finds
\begin{equation}
\label{res0}
\partial^2_{t\omega}Ê\mL-\partial^2_{xk}Ê\mL=\int \mH_t \partial_x f_t dI.
\end{equation}
As mentioned above, in $\mL$, we account for the contribution from electrons whose orbit in phase space may be arbitrarily close to the separatrix. However, the effect of separatrix crossing must be treated with care, for reasons that will be detailed in Paragraph~\ref{III.2.2}. Hence, strictly speaking, \reso~only holds when trapping or detrapping may be ignored, which is true when the density of the electrons that cross the separatrix is negligible. Moreover,~\reso~needs to be solved together with the consistency relation, $\partial_t k=-\partial_x\omega$, and with the Lagrange equations for the other fields, namely, the potential amplitudes $\Phi_n$ or $\delta A$. Let $\mF$ be any of these fields, then, as shown in Appendix~\ref{C},
\begin{equation}
\label{36}
\frac{\partial}{\partial x}\left[\frac{\partial \mL}{\partial \mF_x}Ê\right]+\frac{\partial}{\partial t}\left[\frac{\partial \mL}{\partial \mF_t}Ê\right]=\frac{\partial \mL}{\partial \mF},
\end{equation}
where, for $w=x$ or $w=t$, $\mF_w \equiv \partial_w \mF$. Note that, except when $\mF=\Phi_0$ or $\mF=\delta A$, Eq.~(\ref{36}) just reads, $\partial_{\mF}\mL=0$. Using Lagrange equations for the potential amplitudes yields the relative amplitudes of the Fourier harmonics, in $\psi$, of the electric field. When these are expressed in terms of the first one, $E_1$, one may  derive the nonlinear dispersion relation, $\omega \equiv \omega(k,E_1)$. Such a procedure has been used in Ref.~\cite{lindberg}, and will not be addressed here. Once the relative amplitudes of the harmonics are known, the electric field may be written as $E(x,t) =E_1(x,t) S(\psi,x,t)$, where $S$ is some 2$\pi$-periodic function in $\psi$. Then, provided that $S$ does not vary too much is space and time, \reso~may be viewed as an equation for $E_1$, whose evolution is continuously described from the linear regime, using Eq.~(\ref{res1}), to the nonlinear one. 

Before concluding this Subsection, let us note that, if the wave is driven by a sinusoidal electric field, $E_d \cos(\psi+\delta \psi)$, then, as shown in Appendix~\ref{D}, one needs to add the term $-\varepsilon_0 E_1 E_d \cos(\delta \psi)/2$ to the right-hand of~\reso. This generalizes the linear result, \resun. 

\subsubsection{Envelope equation when trapping and detrapping are allowed}
\label{III.2.2}
Although~\reso~is valid even when some electron orbits are very close to the separatrix, the effect of separatrix crossing is not accounted for in this equation. Hence,~\reso~Êactually applies to the following physics siuation. The EPW has grown from noise and has trapped particles up to a given time $t_0$, while, up to the time $t_0+\Delta t$, a negligible amount of electrons is either trapped or detrapped. Then,~\reso~is only valid to describe the EPW propagation during the time interval $(t_0,t_0+\Delta t)$. 

Now, in order to allow for trapping or detrapping, one needs to define the Lagrangian density so that it would remain continuous through seperatrix crossing, which requires a continuous definition for the dynamical action, as that introduced in Appendix~\ref{A}. From this definition, it follows that the action of a given trapped electron depends on the value assumed by $uÊ\equiv v_\phi-eA/m$ at the very time when trapping occurred. Consequently, the Lagrangian density of the trapped electrons, at time $t$, depends on all the values of $u$ assumed at times $t'<t$. Hence, $\mL_t$ non locally depends on $u$, which is one difficulty to use Lagrange equations. However, if one is only interested in the EPW propagation during the time interval $(t_0,t_0+\Delta t)$, when one may assume that no electron crosses the separarix, then, all the values assumed by $u$ for times $t'<t_0$ may be considered as given, constant, parameters. Then, the Lagrangian density may, again, be considered as local, which explains why, \textit{a priori},~\reso~is only valid during the time interval $(t_0,t_0+\Delta t)$. Moreover, this equation was derived by assuming that $f_t$ remained constant in the wave frame, while this is no longer true when continuous separatrix crossing is allowed. However, the aforementioned difficulties are easily overcome, as we now show it.

Let us first denote, $\mL_\omega \equiv \partial_\omega \mL$. Then, if during the time interval $(t_0,t_0+\delta t)$, a negligible amount of electrons crosses the separatrix, $\partial _t \mL_\omegaÊ\equiv \partial_t \mL_\omega^{(0)}$, where $\partial_t \mL_\omega^{(0)}$ is the time variation of $\mL_\omega$ derived from~\reso. If, now, the contribution from the electrons that are either being trapped or detrapped may no longer be neglected,~\reso~is changed into
\begin{equation}
\label{47}
\partial_t \mL_\omega=\partial_t \mL_\omega^{(0)}+\delta t \mL_{t_0}^{(S)},
\end{equation}
where the term $\delta t\mL_{t_0}^{(S)}$ accounts for the effect of separatrix crossing, and is still unknown. Now, clearly, this term is proportional to the density, $n^{(S)}$, of electrons which cross the separatrix between $t_0$ and $t_0+\delta t$, so that $\mL_{t_0}^{(S)}Ê\propto n^{(S)}/\delta t$.  Now, from the conditions Eqs.~(\ref{u})~and~(\ref{t}) for trapping and detrapping, one finds that, in the limit when $\delta t \rightarrow 0$, $n^{(S)}/\delta t \propto d\mA_s/dt$ (where $\mA_s$ is the area of the separatrix), and remains finite as long as $\mA_s$ remains away from zero. Here, we restrict to the nonlinear situation when the electron response to the wave is adiabatic. Hence, $\mA_s$ is bounded from below, and $\mL_{t_0}^{(S)}$ remains finite as $\delta t \rightarrow 0$. Then, in order to find the evolution of $\mL_\omega$ between $t=0$ and any time $t_0$, one may divide the time interval $(0, t_0)$ into $N$ time intervals of amplitude $\delta t$ to find, from Eq.~(\ref{47}),
\begin{equation}
\label{48}
\mL_\omega(t_0)-\mL_\omega(0)=  \sum_{n=1}^{N}Ê\int_{(n-1)\delta t}^{n \delta t} \partial_t \mL_\omega^{(0)} dt +\delta t \sum_{n=1}^{N}Ê\int_{(n-1)\delta t}^{n \delta t} \mL_{n\delta t}^{(S)} dt.
\end{equation}       
We may now take the limit $\delta t \rightarrow 0$, and, since $ \mL_{n\delta t}^{(S)}$ remains finite in this limit, the last term in the right-hand side of Eq.~(\ref{48}) vanishes, so that one is left with
\begin{equation}
\label{49}
\mL_\omega(t_0)-\mL_\omega(0)= \int_0^{t_0} \partial_t \mL_\omega^{(0)} dt,
\end{equation}
which is, formally, equivalent to~\reso. However, there are caveats in directly using \reso, which we now explain. First, in this equation, the range in action of the trapped and passing electrons is supposed to be constant, and must therefore be considered as such when calculating the derivatives of $\mL$, even though they change due to separatrix crossing. Moreover, as regards $f_u(X,P,t)$, in~\reso, it is supposed to be known at a given time, $t_0$, and to evolve according to  Eq.~(\ref{30}) when $t>t_0$. Now, if some electrons are detrapped at $t=t_{detrap}$, detrapping occurs symmetrically with respect to the phase velocity~\cite{dissipation,comment4}, which entails an effective change in $f_u$. However, in the envelope equation, one must not account for this change in $f_u$ when calculating the derivatives of $\mL$. In these derivatives must only enter the evolution of $f_u$ when $t>t_{detrap}$, given by Eq.~(\ref{30}). Note that the change in $f_u$ due to symmetric detrapping simply follows from a geometric change in the action, by the width of the separatrix, which is not in contradiction with the adiabatic evolution of the electrons~\cite{ratchet}.

Due to trapping and detrapping, $\mL$ becomes an explicit function of the area of the separatrix~$\mA_s$ (and this dependence is even non local for $f_u$). Then, in order to avoid any confusion regarding the actual meaning of \reso~when separatrix crossing is allowed, we change this equation into,
\begin{equation}
\label{51}
\partial_t \mL_\omega \vert_{\mA_s}-\partial_x \mL_k \vert_{\mA_s}=\int \mH_t \partial_x f_tdI,
\end{equation}
in order to specify that, when calculating the derivatives of $\mL$, its variations entailed only by trapping and detrapping (and, therefore, by a change in $\mA_s$), must not be accounted for. Hence, in Eq.~(\ref{51}), the space and time derivatives are not full derivatives, so that, even if the term in the right-hand side of Eq.~(\ref{51}) was zero, $\int \mL_\omega dx$ would not be conserved. This is due to the dissipation entailed by trapping. Indeed, as discussed in Refs.~\cite{vgroup,dissipation}, trapping entails an increase in the electron kinetic energy, which is not restored when they are detrapped, because detrapping occurs symmetrically with respect to the phase velocity. Hence, trapping entails an irreversible increase in kinetic energy and is, therefore, an effective means of dissipation for the EPW. As shown in Refs.~\cite{vgroup,dissipation}, this dissipation leads to no damping of the wave packet, but to the decrease of its space extent. Actually, it is easily checked that, when the wave number and frequency are constant,  and when the wave is not driven, Eq.~(\ref{51}) just amounts to Eq.~(22) of Ref.~\cite{vgroup}, although these equations have been derived in a completely different way. 

\section{Connection between perturbative and near-adiabatic results}
\label{new}
In Section~\ref{III}, we derived two different envelope equations for the EPW, obtained by assuming two different electron responses to the wave, a near-adiabatic and a linear one. These two responses are usually thought of as corresponding to two completely different regimes of wave-particle interaction. Indeed, linear results are expected to be accurate for vanishingly small amplitudes. By contrast, the near-adiabatic ones are valid once the bounce period, $T_B$, is much smaller than any other timescale, which only occurs for large enough wave amplitudes (since, for a sinusoidal wave, $T_B$ scales as $1/\sqrt{E_1}$). However, we show in this Section that, by simply connecting the results from these two theories, one may obtain a very accurate envelope equation, valid whatever the regime of wave-particle interaction (and the accuracy may even be increased by replacing the linear results with those derived from a high order perturbation analysis, in the wave amplitude). To prove the latter assertion we first address, in Paragraph~\ref{new1}, the small amplitude limit of the adiabatic formulas, and discuss how they may match the linear  results. Based on this discussion, we explain, in Paragraph~\ref{new2}, how to actually make the connection between the linear (or perturbative) wave equation and the near-adiabatic one in the homogeneous case (while the general inhomogeneous situation is addressed in Section~\ref{IV}). The accuracy of the envelope equation thus obtained is then tested against numerical simulations in Paragraph~\ref{new3}. 

\subsection{Small amplitude limit of the near-adiabatic formulas}
\label{new1}
Since the limit of small amplitudes refers to nearly sinusoidal waves, only such waves will be considered in this Paragraph. Moreover, we restrict to the situation when the phase velocity varies more slowly than the trapping velocity, which translates to, $\vert dv_\phi/dt \vert < (4/\pi) \vert d(\omega_B/k)/dt \vert$.  Under this condition, the electrons that cross the separatrix are always trapped in the wave trough. Hence, we only consider the situation when the effect of trapping is maximum. Indeed, if, at a given time,  $\vert dv_\phi/dt \vert > (4/\pi) \vert d(\omega_B/k)/dt \vert$, the electrons crossing the separatrix from above (respectively from below) would remain untrapped, below (respectively above) the separatrix~\cite{ratchet}. Then, the populations of trapped and untrapped electrons would depend on the whole history of the wave evolution, and each particular situation would need to be studied individually. Therefore, for a general description, one needs to restrict to $\vert dv_\phi/dt \vert < (4/\pi) \vert d(\omega_B/k)/dt \vert$, although any different situation could be addressed along similar lines. 

For a sinusoidal EPW, the electric field reads, $E=E_1\sin(\psi)$, and we use $E_1$, $k$, $\omega$ and $\psi$ as the independent variables of the adiabatic Rouhtian density, $\mL$, derived in Section~\ref{III}. Moreover, in this Paragraph, we only investigate the derivatives of $\mL$ with respect to $E_1$ or $\omega$. Indeed, only these derivatives are useful to our discussion, and may lead to a direct comparison with the results from the test particles simulations of Paragraph~\ref{new2}. \\

From Eq.~(\ref{E10}), the adiabatic dispersion relation for a sinusoidal wave is,
\begin{eqnarray}
\nonumber
1+\frac{2e}{\varepsilon_0 k E_1}&&\left\{\int_{\frac{4\omega_B}{\pi k}}^{+\infty} \left[1+\frac{2}{\zeta_u}\left(\frac{K_2}{K_1}-1 \right)Ê\right]\delta_+F_u(V_u)dV_uÊ\right.\\
\nonumber
&&\left.+\int_{0}^{\frac{4\omega_B}{\pi k}}\left[\frac{2K_2}{K_1}-1\right]\delta_+F_t(V_t)dV_t+\frac{8}{3\pi}Ê\left[F_u(v_\phi)-F_t(v_\phi)Ê\right] Ê\frac{\omega_B}{k}\right\}=0,\\
\label{n1}
\end{eqnarray}
where $\omega_B=\sqrt{eE_1k/m}$, where the variables $V_u$ and $V_t$ are, respectively, defined by Eq.~(\ref{E00})~and~(\ref{E01}), and where the distribution functions $F_u(V_u)$ and $F_t(V_t)$ are such that $F_u(v_\phi+V_u)dV_u=f_u(P)dP$ and $F_t(v_\phi+V_t)dV_t=f_t(I)kdI$. $K_1$ and $K_2$ are, respectively, the Jacobian elliptic integrals of first and second kind whose arguments are defined the following way. In the first integral of the left-hand side of Eq.~(\ref{n1}), the argument, $\zeta_u$, is related to $V_u$ by Eq.~(\ref{E00}), while in the second integral of the left-hand side Eq.~(\ref{n1}), the argument, $\zeta_t$,  is related to  $V_t$ by Eq.~(\ref{E01}). Finally,  $\delta_+F(V) \equiv F(v_\phi+V)+F(v_\phi-V)-2F(v_\phi)$. Hence, clearly, the second term in the brackets of Eq.~(\ref{n1}) is of the order of $(\omega_B/k)^{3}$ when $E_1 \rightarrow 0$. Moreover, using Eq.~(\ref{E00}), it is easily shown that, $C(V_u) \equiv 1+2(K_2/K_1-1)/\zeta_u$, is such that $C(V_u)=-eE_1/2mkV_u^2+O(e^2E_1^2/k^2m^2V_u^4)$ when $\omega_B/kV_u \rightarrow 0$. Then, let us introduce some arbitrary amplitude, $E_A$, some dimensionless number, $\iota$, such that $0<\iota<1/2$, and, when $E_1<E_A$, let us decompose the first integral in the left-hand side of Eq.~(\ref{n1}) the following way,
\begin{equation}
\label{n2}
\int_{\frac{4\omega_B}{\pi k}}^{+\infty}C(V_u)\delta_+F_u(V_u)dV_uÊ= \left[\int_{\frac{4\omega_B}{\pi k}}^{\frac{4\omega_B}{\pi k}\left(\frac{E_A}{E_1}\right)^\iota }Ê+\int_{\frac{4\omega_B}{\pi k}\left(\frac{E_A}{E_1}\right)^\iota }^{+\infty} \right]C(V_u)\delta_+F_u(V_u)dV_u.
\end{equation}
In the second integral of the right-hand side of Eq.~(\ref{n2}), $\omega_B/kV_u <\pi (E_1/E_A)^\iota/4$, and goes to zero when $E_1$ vanishes, so that this integral converges towards $-(eE_1/2mk) P.P. \left(\int \delta_+ F_u(V_u)/V_u^2 dV_u \right)$. In the first integral in the right-hand side of Eq.~(\ref{n2}), when $E_1$ vanishes, $V_u$ goes to zero so that $\delta_+F_u(V_u)=O(V_u^2)$. Hence, this integral is $O\left[(eE_1/mk)^{3/2}(E_A/E_1)^\iota\right]$ and, therefore, negligible compared to first one. We conclude that,
\begin{eqnarray}
\nonumber
\lim_{E_1 \rightarrow 0} \int_{\frac{4\omega_B}{\pi k}}^{+\infty}C(V_u)\delta_+F_u(V_u)dV_uÊ& = & \frac{-eE_1}{2mk} P.P. \left(\int \frac{\delta_+ F_u(V_u)}{V_u^2} dV_u \right)\\
\label{n3}
&=& \frac{\varepsilon_0 E_1 k}{2e}Ê\chi_{lin},
\end{eqnarray}
where $\chi_{lin}$ is defined by Eq.~(\ref{20})~\cite{note}. Hence, in the small amplitude limit, the nonlinear adiabatic dispersion relation reads,
\begin{equation}
\label{n4}
1+\chi_{lin}+\frac{16 \omega_{pe}^2}{3\pi k \omega_B}Ê Ê\frac{F_u(v_\phi)-F_t(v_\phi)}{n}=0,
\end{equation}
where $n$ is the local electron density. Clearly, only when $F_u(v_\phi)=F_t(v_\phi)$ does Eq.~(\ref{n4}) make sense in the limit $E_1 \rightarrow 0$ , in which case one recovers the linear dispersion relation, $1+\chi_{lin}=0$. 

Let us now comment the condition, $F_u(v_\phi)=F_t(v_\phi)$, we have to impose in order to recover the linear limit, which actually amounts to discussing the very notion of trapping.  When the wave amplitude is constant, or when making use of the adiabatic approximation, electrons are considered as trapped when $\zeta_u \equiv 2\mh/(\mh+\Phi_1)$ is less than unity. However, for a growing wave, it has been proved in Ref.~\cite{action} that a perturbation analysis, in the wave amplitude, may very accurately predict the locations in phase space of electrons such that $\zeta_u \alt 1$. Hence, when it comes to deriving their distribution function, such electrons must clearly not be considered as trapped, since the very notion of trapping is irrelevant in a perturbation analysis. Actually,  trapping is only effective once the most resonant electrons have completed a nearly closed orbit, i.e., when $\int \omega_B dt \agt 2\pi$. Only then does Eq.~(\ref{A6b}) apply to calculate  the trapped distribution function. More generally, as discussed in several papers (see Ref.~\cite{srs3D}~and references therein), and in Paragraphs~\ref{new2}~and~\ref{new3}, only when $\int \omega_B dt \agt 5$ are adiabatic results expected to be accurate. However, Eq.~(\ref{n4}) shows that the adiabatic dispersion relation remains valid in the limit $E_1 \rightarrow 0$, provided $F_t$ be defined so that $F_t(v_\phi)=F_u(v_\phi)$. Note that the latter equality is always satisfied if the plasma is homogeneous. In a more general situation, the trapped distribution function must be considered as somewhat ``enslaved'' to the untrapped, i.e., when $\int \omega_B dt \alt 5$, the density of trapped electrons must be related to that of the untrapped in order for the distribution function to remain continuous through separatrix crossing.  Hence, we conclude that the adiabatic dispersion relation is always valid provided that $F_t$ be correctly defined. When $\int \omega_B dt \alt 5$, it must be calculated as if the plasma was homogenous, with a density related to that of the untrapped, while, when $\int \omega_B dt \agt 5$,  it must obey Eq.~(\ref{A6b}). Hence, there is some arbitrariness is the definition of $F_t$, regarding the value of $\int \omega_B dt$ beyond which Eq.~(\ref{A6b}) must be used.  Nevertheless, this does not  affect the dispersion relation if the plasma inhomogeneity is small enough. More precisely, as will be discussed in Paragraphs~\ref{new2}~and~\ref{new3}, the transition between the perturbative and adiabatic regimes of propagation occurs within the time $\delta t$ during which $\int \omega_B dt$ varies from about 4.5 to about 5. If, during this time, and within the $x$-extent $v_\phi \delta t$, the density varies very little, $F_t$ would be essentially the same whether it is derived from Eq.~(\ref{A6b}) or as if the plasma was homogeneous. \\

Let us now address the small amplitude limit of $E_1^{-1}\partial^2_{E_1\omega} \mL$, which yields the term proportional to $\partial_t E_1$ in the envelope equation. From Eq.~(\ref{E13}),
\begin{equation}
\label{n5}
-\frac{1}{E_1}\frac{\partial^2 \mL}{\partial E_1\partial \omega}=-\frac{e}{k^2E_1} \int^{+\infty}_{\frac{4\omega_B}{\pi k}} \delta_-F_u(V_u)\frac{\partial C(V_u)}{\partial V_u} dV_u-\frac{32e^2}{3\pi m k^2\omega_B}F'_u(v_\phi),
\end{equation}
where $\delta_-F(V) \equiv F(v_\phi+V)-F(v_\phi-V)-2VF'(V)$, and $F'(V) \equiv \partial_{V}F$. Now, using the same technique as for the dispersion relation, it is easily shown that the first term in the right-hand side of Eq.~(\ref{n5})~converges towards $(\varepsilon_0/2)\partial_\omega \chi_{lin}$  when $E_1 \rightarrow 0$. Hence, for small values of $E_1$,
\begin{equation}
\label{n6}
E_1^{-1}\partial_{E_1 \omega} \mLÊ\approx \frac{\varepsilon_0}{2}Ê\partial_\omegaÊ\chi_{lin}\left[1+\frac{64}{3\pi^2} \frac{\nu_L}{\omega_B}Ê\right],
\end{equation}
where we have used  the expression~(\ref{21})~for the Landau damping rate, $\nu_L$ \cite{note}. Therefore,  in the small amplitude limit, the adiabatic wave equation will contain the following terms,
\begin{equation}
\label{n7}
E_1Ê\partial_\omegaÊ\chi_{lin} \left[\partial_t E_1+\frac{64}{3\pi^2} \frac{\gamma}{\omega_B}\nu_LÊE_1\right],
\end{equation}
where $\gamma \equiv E_1^{-1}\partial_t E_1$ is the wave growth rate. From Eq.~(\ref{res1}), the linear counterpart of these terms is, $E_1\partial_\omegaÊ\chi_{lin}\left(\partial_t E_1 +\nu_L E_1Ê\right)$. At first sight, the linear limit is not recovered since, in the adiabatic regime, the linear Landau damping rate is replaced by a differential operator. However, the \textit{numerical} value assumed by the sum of the terms in Eq.~(\ref{n7})~matches the linear one when $\omega_B/\gamma = 63/3\pi^2 \approx 2.15$. For a growing wave, this condition translates into $\int \omega_B dt \approx 4.3$, a value close to $2\pi$. This is not by chance, and this has a very deep physical meaning that we now explain. 

As is well known (see, for example, Ref.~\cite{DNC}), Landau damping cannot be recovered in the adiabatic limit because, in this limit, the electron distribution is assumed to be phase mixed (i.e., independent of the angle, $\theta$). Moreover, as shown in Ref.~\cite{dissipation}, due to the very same phase mixing, trapping entails an irreversible increase of the electron kinetic energy. Therefore, it is an effective means of dissipation for the wave, which does not entail any damping but the shrinking of the wave packet, as discussed in Paragraph~\ref{III.2.2}. The fact that Eq.~(\ref{n7})~matches the linear value when $\int \omega_B dt \approx 2\pi$ just means that, when trapping becomes effective, although dissipation changes from Landau damping to the reduction of the extent of the wave packet, the wave growth rate remains continuous, as should be. 
Now, mathematically speaking, due to phase mixing, the contribution from the trapped electrons to the adiabatic Routhian density, $\mL$, is independent of $\omega$. Consequently,  unlike for the dispersion relation, there is no counterpart from the trapped electrons to the last term in the right-hand side of Eq.~(\ref{n6}),~that may cancel it out. This term, responsible for the dissipation entailed by trapping, has to remain, and its numerical value is shown to effectively match the Landau damping rate. The latter results, discussed in previous publications, Refs.~\cite{vgroup,dissipation}, is recovered here in a very simple and elegant fashion, thanks to the variational formalism. \\

Finally, let us address the small amplitude limit of $E_1^{-2}\partial_\omega^2 \mL$, which yields the term proportional to $\partial_t \omega$ in the envelope equation. From Eq.~(\ref{E16}),
\begin{eqnarray}
\nonumber
-\frac{1}{E_1^2}\frac{\partial^2\mL}{\partial \omega^2}=&&-\frac{e^2}{k^4E_1^2}\int^{+\infty}_{\frac{4\omega_B}{\pi k}} \delta^2_+F_u(V_u)\frac{\partial^2\mH_u}{\partial V_u^2}dV_u\\
\nonumber
&&+\frac{me^2}{E_1^2}\int_{-\frac{4\omega_B}{\pi k}}^{\frac{4\omega_B}{\pi k}}\frac{F_t(v_\phi+V)-F_u(v_\phi)-(V^2/2)F_u''(v_\phi)}{k^4}dV \\
\label{n8}
&&-\frac{64e^2}{3\pi^2k^3m\omega_B}F''_u(v_\phi),
\end{eqnarray}
where $\delta^2_+F(V) \equiv F(v_\phi+V)+F(v_\phi-V)-2F(v_\phi)-V^2F_u''(v_\phi)$. Using the same kind of calculation as for $\partial_{E_1} \mL$, it is easily shown that the first term in the right-hand side of Eq.~(\ref{n8}) converges towards $(\varepsilon_0/4) \partial_{\omega^2}\chi_{lin}$. Moreover, as already discussed for the nonlinear dispersion relation, in the small amplitude limit, one must use $F_u=F_t$. Then, it is clear that the second term in the right-hand side of Eq.~(\ref{n8}) scales as $\sqrt{E_1}$ and is negligible in the limit $E_1 \rightarrow 0$. Hence, for small amplitudes, the adiabatic wave equation contains the following terms
\begin{equation}
\label{n9}
E_1^{2} \left[\frac{\partial_{\omega^2}Ê\chi_{lin}}{2}-\frac{128 \omega_{pe}^2}{3\pi n k^3\omega_B} F''_u(v_\phi)\right]\partial_t \omega.
\end{equation}
Now, from Eq.~(\ref{res1}),  only the term $E_1^{2}(\partial_{\omega^2}Ê\chi_{lin}/2)\partial_t \omega$ is present  in the linear envelope equation. Therefore, unlike for the dispersion relation and for collisionless dissipation, we find a new term in the adiabatic wave equation, namely the last term in the bracket of Eq.~(\ref{n9}), which has no  linear counterpart. Moreover, in the small amplitude limit, this term is dominant in Eq.~(\ref{n9}). Let us discuss it physically. The averaged velocity of trapped electrons is the wave phase velocity. Hence, when it changes, for example, due to a change in the wave frequency, it entails a change in the average kinetic energy of the trapped electrons, at the expense (or benefit) of the EPW. The second term in the bracket of Eq.~(\ref{n9}) accounts for this effect, which is directly entailed by trapping, and, therefore, has no linear counterpart. 
Note that, although the density of trapped electrons scales as $\sqrt{E_1}$, their acceleration due to a change in the phase velocity does not lead to a term, in the wave equation, scaling the same way, unless what has been inferred in Refs.~\cite{dodinIII,schmit}. Indeed, we find here that there is a subtle cancellation between the contributions from the trapped and untrapped electrons, that makes this term rather scale as $E_1^{3/2}$.  Note, also, that Eq.~(\ref{n9}) has been obtained by assuming that the wave phase velocity varies less rapidly than its trapping width. When the wavenumber is constant, this translates into $\vert \partial_t\omega \vert <(2\gamma\omega_B/\pi )$, so that the terms in Eq.~(\ref{n9}) are of lower order than those in Eq.~(\ref{n7}). This insures some continuity in the wave equation when shifting  from the linear to the near-adiabatic regime, as should be. 

\subsection{Connection between perturbative and near-adiabatic results in the homogeneous case}
\label{new2}

Linear theory actually stems from a first order analysis, in the wave amplitude, of the electron motion. Now, such an analysis may be performed to higher order, as explained in Ref.~\cite{benisti07}. In short, starting from Eqs.~(\ref{A7})-(\ref{A8}) for the Hamiltonian ruling the electron dynamics, the perturbation analysis consists in introducing a new set of conjugated variables such that, in these new variables, the scalar potential, $\Phi$, may be neglected up to a very small term, of order $\varepsilon^l$. Here, $l$ is the order up to which the analysis has been led, while, for a wave growing at the rate $\gamma$, the small parameter of the analysis, $\varepsilon$, was found in Ref.~\cite{benisti07} to be $\varepsilon = \omega_B^2/[(\gamma/k)^2+(v_0-v_\phi)^2]$, $v_0$ being the initial electron velocity (when the wave amplitude is vanishingly small). Such a formula for $\varepsilon$ makes sense because it is clear that nearly resonant electrons, whose speed is close to the phase velocity, are most effectively affected by the wave while, if the wave amplitude changes at rate $\gamma$, the resonance is blurred over a velocity width of the order of $\gamma/k$. The value found for $\varepsilon$ shows that, as the wave amplitude increases, the motion of fewer electrons may be accurately derived perturbatively. More precisely, in Ref.~\cite{benisti07} it was found that, for a slowly growing sinusoidal wave, perturbative results only apply to electrons such that,
\begin{equation}
\label{n10}
\vert v_0-v_\phi\vert > V_{lim},
\end{equation}
\begin{equation}
\label{n10b}
V_{lim}=\max[0,\ (4\omega_B/\pi k)(1-3\gamma/2\omega_B)]. 
\end{equation}
Since, for a growing wave, $2\omega_B/\gamma \approx \int \omega_B dt$, Eqs.~(\ref{n10})~and~(\ref{n10b})~just mean that perturbative results are no longer valid when the electrons have completed about one half of a trapped orbit, which makes sense. Moreover, only in the perturbative regime may the electrons effectively contribute to Landau damping, so that~Eqs.~(\ref{n10})~and (\ref{n10b})~actually provide a way to calculate the first nonlinear correction to the Landau damping rate. More precisely, using the results of Ref.~\cite{benisti07}, one finds that, for a homogeneous growing wave, a first order perturbation analysis, together with the criterion~(\ref{n10}), leads to the following envelope equation, 
\begin{equation}
\label{n11}
\partial_\omega \chi_{1}Ê\left[\partial_t E_1+\nu_1ÊE_1 \right]+(E_1/2) \partial_{\omega^2}\chi_{1}Ê\partial_t \omega = E_d,
\end{equation}
where $E_d$ is some drive amplitude [that may be zero, and to which the dephasing term, $\cos(\delta \psi)$, has been added]. Moreover, in Eq.~(\ref{n11}), 
\begin{equation}
\label{n12}
\chi_1= -\frac{\omega_{pe}^2}{nk} \int_{\vert v_0-v_\phi\vert\leq V_{lim}} \frac{f^{'}_H}{kv_k-\omega}dv_k,
\end{equation}
where the integral in Eq.~(\ref{n12}) has to be understood as its principal part when $V_{lim}=0$, and
\begin{equation}
\label{n12}
\nu_1 \partial_\omega \chi_{1} =-\frac{\omega_{pe}^2f'_H(v_\phi)}{nk^2} \left[Ê\pi-2\tan^{-1}\left(\frac{k V_{lim}}{\gamma}\right) + \frac{2 \gamma k V_{lim}}{\gamma^2+(kV_{lim})^2}Ê\right],
\end{equation} 
where $f_H$ is the unperturbed distribution function, as defined in Paragraph~\ref{III.1}. Note that, when $V_{lim}=0$, $\chi_1=\chi_{lin}$, and $\nu_1$ is nothing but the linear Landau damping rate.  Moreover, when $V_{lim}Ê\gg \gamma/k$, $\nu_1$ is nearly proportional to $\gamma$, so that one recovers a result akin to the near-adiabatic one, Eq.~(\ref{n7}), i.e., the damping rate has nearly vanished and has been replaced by a term proportional to the growth rate. However, for this particular term, using a perturbation analysis, one would find that the factor $64/3\pi^2$ in Eq.~(\ref{n7}) would be replaced by  unity. This shows the limit of a first order perturbation analysis, it cannot be used for large amplitudes, such that $\omega_BÊ\gg \gamma$. Now, clearly, the results obtained at first order, and the near-adiabatic ones, have to be connected when they are closer to each other, i.e., when the term $(64/3\pi^2)(\gamma/\omega_B) \nu_L$ is closer to $\nu_1$. This is easily found to be when $\omega_B/\gamma \approx 2.3$, in which case both terms are quite close to each other and differ by less than 15\%. Hence, one may just shift abruptly from the first-order to the near-adiabatic envelope equation, so that these equations are connected using a Heaviside-like function. Namely, in the homogeneous case, our wave equation would read,
\begin{equation}
\label{n12}
\varepsilon_0E_dE_1/2 = (1-Y_1)\left[\partial_{\omega E_1} \Lambda_1Ê\partial_t E_1+\partial_{\omega^2} \Lambda_1Ê\partial_t \omega+ 2\nu_1\partial_\omega \Lambda_1\right] - Y_1\left[\partial_{\omega E_1} \mLÊ\partial_t E_1+\partial_{\omega^2} \mLÊ\partial_t \omega\right],
\end{equation}
where we have denoted $\Lambda_1 \equiv \varepsilon_0 \chi_{1}E_1^2/4$, and where we have chosen
\begin{equation}
\label{n13}
Y_1=\tanh^8\left[\left(e^{\omega_B/3\gamma}-1\right)^3\right].
\end{equation}
$Y_1 \approx 0.5$ when the first order and near-adiabatic results nearly match, i.e., when $\omega_B/\gamma \approx 2.3$. Moreover, $Y_1$ varies very quickly from zero to unity about $\omega_B/\gamma=2.3$. More precisely, $Y_1<0.1$ when $\omega_B/\gamma \alt 2.05$ and $Y_1>0.9$ when $\omega_B/\gamma \agt 2.6$. Therefore, Eq.~(\ref{n12}) amounts to the perturbative wave equation when $\int \omega_B dt \alt 4$ and to the near-adiabatic one when $\int \omega_B dt \agt 5$, which is a little less than the intuitive value $\int \omega_B dt \agt 2\pi$. For the sake of simplicity, we omitted to specify that, for the near-adiabatic part of the wave equation, the derivatives need to be calculated for a fixed value of $\mA_s$. Such a specification is actually not compulsory if the wave keeps growing, because detrapping never occurs.  As shown numerically in Paragraph~\ref{new3}, the wave equation~(\ref{n12}) is quite accurate. 

The accuracy may even be increased by using a higher order perturbation analysis, which we did up to order 11. Then, the wave equation~(\ref{n12})~is changed into
\begin{equation}
\label{n14}
\varepsilon_0E_dE_1/2 = (1-Y_{11})\left[\partial_{\omega E_1} \Lambda_{11}Ê\partial_t E_1+\partial_{\omega^2} \Lambda_{11}Ê\partial_t \omega+ 2\nu_{11}\partial_\omega \Lambda_{11}\right] - Y_{11}\left[\partial_{\omega E_1} \mLÊ\partial_t E_1+\partial_{\omega^2} \mLÊ\partial_t \omega\right],
\end{equation}
where $\Lambda_{11}$ and $\nu_{11}$ are the eleventh order counterpart of $\Lambda_1$ and $\nu_1$ and where, now, we choose, 
\begin{equation}
\label{n15}
Y_{11}=\tanh^{80}\left[\left(e^{\omega_B/2.8\gamma}-1\right)^3\right].
\end{equation}
$Y_{11}\approx 0.5$ when $\omega_B/\gamma \approx 2.44$, a value a little bit higher than for $Y_1$ since results at higher order are accurate up to larger amplitudes. Moreover, since the high order perturbation results almost perfectly match the near-adiabatic ones, $Y_{11}$ changes much more rapidly from 0 to unity than $Y_1$. In particular, $Y_{11}Ê<.1$ when $\omega_B/\gamma \alt 2.3$ and $Y_{11}Ê>.9$ when $\omega_B/\gamma \agt 2.6$, so that Eq.~(\ref{n14}) amounts to the perturbative wave equation when $\int \omega_B dt \alt 4.5$ and the near-adiabatic one when $\int \omega_B dt \agt 5$. Hence, the conclusions drawn at first order do not change much when using the 11th order, especially as regards the validity of the near-adiabatic approximation. As regards $\Lambda_{11}$ and $\nu_{11}$,  their expressions are rather lengthy and will not be given here, but may be easily inferred from the results of Ref.~\cite{benisti07}. Moreover, it is important to note that, both $\nu_{11}$ and $\nu_1$ remain quite close to the Landau damping rate when $Y_{11}$, or $Y_1$, is close to zero. Consequently, in the wave equation~(\ref{n12}), one may replace $\nu_1$ by $\nu_L$, and this approximation will be systematically used in Section~\ref{IV}. Similarly, $\chi_1 \approx \chi_{lin}$, so that $\Lambda_1$ may be replaced by $\Lambda_{lin}$. \\

So far, the connection of the perturbative and near-adiabatic wave equations has only been adressed for a growing wave. Hence, the considered situation is that of a wave that grows from the noise level and keeps growing when entering the nonlinear, near-adiabatic, regime (although its evolution, after reaching the near-adiabatic regime, may be arbitrary). This situation is quite general since, as discussed above, the transition from the perturbative to the near-adiabatic regimes is quite abrupt. Using the hypothesis of a growing wave, we could provide an explicit expression for $\nu_1$, which actually made Eq.~(\ref{n11})~be more an algebraic equation for $\gamma$ than a differential equation. However, results valid whatever the variations of the wave amplitude may also be obtained. Indeed, as shown in Ref.~\cite{benisti09}, for arbitrary time variations of the wave amplitude, Eq.~(\ref{n11}) generalizes into
\begin{equation}
\label{n15}
\partial_\omega \chi_{lin}Ê\partial_t E_1-\frac{\omega_{pe}^2f'_H}{n}(v_\phi)\int_0^t\int_0^uE_1(\xi)\int_{\vert w \vert \geq V_{lim}} iwe^{ikw(\xi-t)}dwd\xi du +(E_1/2) \partial_{\omega^2}\chi_{lin}Ê\partial_t \omega = E_d,
\end{equation}
where, now,
\begin{equation}
\label{n16}
V_{lim}=\max\left[0,\ (4\omega_B/\pi k)\left(1-3/\int \omega_B dt\right)\right]. 
\end{equation}
Moreover, it is easily shown that, when $V_{lim}=0$, the second term in the left-hand side of Eq.~(\ref{n16}) is just $\partial_\omega \chi_{lin} \nu_L E_1$~\cite{benisti09}. It is also noteworthy that, when $V_{lim} \gg \vert \gamma \vert/k$,  at lowest order in $\gamma/kV_{lim}$, this term matches  the result derived from the expansion of $\nu_1$ for large values of $kV_{lim}/\gamma$. This shows that the envelope equation~(\ref{n11}), derived for a growing wave, bears some relevance whatever the way the wave amplitude varies. 
However, the most accurate way to account for the nonlinear wave evolution in the general situation is not to make use of an effective damping rate, like $\nu_1$, but to solve an integro-differential equation, similar to Eq.~(\ref{n15}). 

Actually, it is possible to describe very precisely the nonlinear modification of Landau damping, even for a wave that does not vary slowly, by making use of a high order perturbation analysis and by explicitly estimating the contribution of the trapped particles along similar lines as in Ref.~\cite{action}. This will be the subject of a forthcoming article, but such a fine description of nonlinear Landau damping is completely outside the scope of this paper. Indeed, here, we aim at deriving a first order envelope equation, much easier to implement in a code that Eq.~(\ref{n15}), and which, nevertheless, provides accurate results. Hence, as discussed above, we simply model the nonlinear Landau damping rate by $(1-Y_1)\nu_L$. Our modeling may not be very accurate when $4.5 \alt \int \omega_B dt \alt 5$ and seems to depend on our choice for $Y_1$. However, as shown in Paragraph~\ref{IV.2}, in a 3-D geometry, the impact of these approximations is very limited as regards our modeling of nonlinear wave propagation, and, in particular, as regards nonlinear Landau damping. 

\subsection{Numerical results}
\label{new3}
 \begin{figure}[!h]
\centerline{\includegraphics[width=13cm]{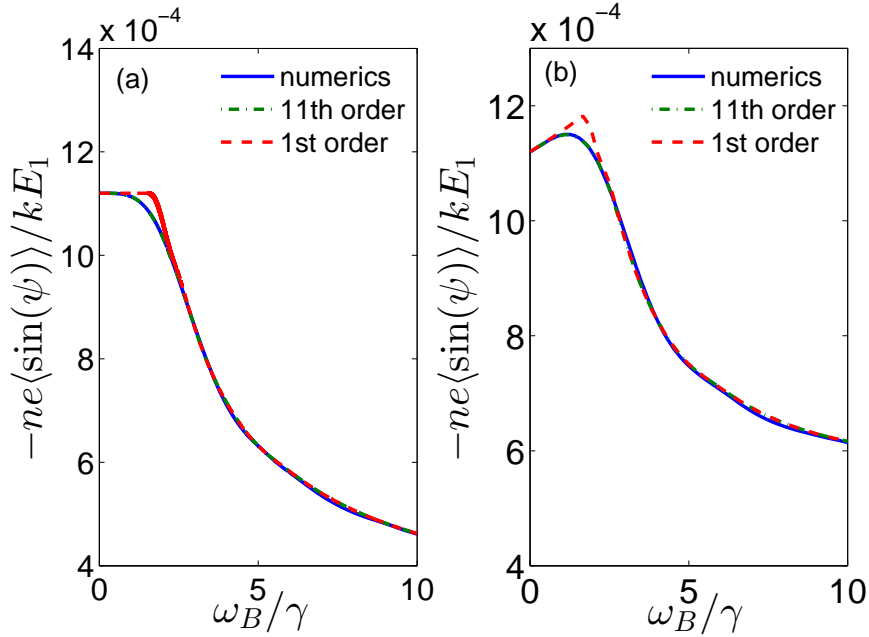}}
\caption{\label{f1}(color online) $-ne \langle \sin(\psi)/kE_1$ as calculated numerically (blue solid line), according to the left-hand side of Eq.~(\ref{n14}) (green dashed-dotted line) and according to the left-hand side of Eq.~(\ref{n12}) (red dashed line), for a wave growing in an initially Maxwellian plasma at the constant rate $\gamma/kv_{th}=10^{-2}$, and whose phase velocity is $v_\phi=4v_{th}$ [panel (a)] or $v_\phi=4v_{th}-\omega_B/k$ [panel (b)].}
\end{figure}
As shown in Ref.~\cite{vlasovia}, for a homogeneous driven sinusoidal wave,
\begin{equation}
\label{n17}
-ne \langle \sin(\psi)/k=\varepsilon_0 E_1E_d/2,
\end{equation}
where the factor $\cos(\delta \psi)$ accounting for dephasing has been included in the definition of $E_d$, and where
\begin{equation}
\label{n18}
\langle \sin(\psi)Ê\rangle(x_0,t) \equiv \frac{1}{2\pi}Ê\int_{\psi(\mathbf{x}_0)-\pi}^{\psi(\mathbf{x}_0)+\pi}\int_{-\infty}^{+\infty} \sin(\psi') f(\psi',v,t)dvd\psi'.
\end{equation}
$\si$ is an averaged value over all electrons located within one wavelength about a given position, just like $\lL$, which explains why we used the same notations for the averaging. Now, for a pointwise distribution function, Eq.~(\ref{n18}) translates into,
\begin{equation}
\label{n19}
\langle \sin(\psi)Ê\rangle(x_0,t)=\frac{1}{N}\sum_{i=1}^N\sin(x_i),
\end{equation}
where the sum is over all electrons located within one wavelength about $x_0$. The identity, Eq.~(\ref{n17}), as simple as it is, provides a very powerful tool to test the accuracy of the envelope equations~(\ref{n12})~or~(\ref{n14}). Indeed, $-ne\langle \sin(\psi) \rangle/k$ is a functional of the electric field. If an envelope equation like Eq.~(\ref{n12})~or~(\ref{n14}) is valid, then, the functional   $-ne\langle \sin(\psi) \rangle/2k$ is nothing but the left-hand side of these equations, \textit{regardless}~of the way the wave amplitude, or wave frequency, varies. In particular, one may use test particles simulations where electrons are acted upon by a \textit{prescribed} electric field, compute numerically $\langle \sin(\psi) \rangle$ according to Eq.~(\ref{n19}), and compare the numerical estimate with the theoretical one, given by the left-hand side of Eq.~(\ref{n12})~or~(\ref{n14}). When the wave frequency is constant, such a technique allows to predict when $\langle \sin(\psi) \rangle$ is proportional to the wave growth rate and, therefore, when the damping rate has become negligible. This procedure has been detailed in Refs.~\cite{benisti07,yampo}, and, in Fig.~\ref{f1}(a), we give one more example illustrating the agreement between the numerical and theoretical values of $\langle \sin(\psi) \rangle$. This Figure is for a wave growing in an initially Maxwellian plasma, at the constant rate, $\gamma=10^{-2}kv_{th}$, while its phase velocity is $v_\phi=4v_{th}$, $v_{th}$ being the thermal velocity. As may be seen in Fig.~\ref{f1}(a), the agreement between theory and numerics is excellent (the relative error is always less than 0.5\%) when making use of a perturbation analysis at order 11, while first order results also show to be quite accurate. The same conclusion holds when the frequency varies, as illustrated in Fig.~\ref{f1}(b) for an electric field which is the same as in Fig.~\ref{f1}(a), except that $v_\phi=4v_{th}-\omega_B/k$, so that the condition $dv_\phi/dt<(4/\pi) d(\omega_B/k)/dt$ is fulfilled. 

The results plotted in Fig.~\ref{f1} call for two comments. First, in Fig.~\ref{f1}(a), when estimating $\si$ perturbatively, only the contribution from electrons such that $\vert v_0 -v_\phi \vert \leq V_{lim}$ has been accounted for. By contrast, the adiabatic estimate of $\si$ is only due to the electrons, such that $\vert v_0 -v_\phi \vert \leq 4\omega_B/k\pi$, considered as ``untrapped'' within the adiabatic approximation. From the definition Eq.~(\ref{n10b})~of $V_{lim}$, the fact that there exists a range in amplitude when the adiabatic and perturbative results do match clearly shows that there exists a range in $v_0$, of the order of $\vert \gamma \vert/k$ (which translates into a range in action of the order of $m\vert \gamma\vert /k^2$), for which the electrons cannot be unambiguously considered as trapped or untrapped. Second, when making use of the adiabatic approximation, the distribution in angle, $\theta$, is assumed to be uniform. Now, as shown in Ref.~\cite{action}, the distribution function in action-angle variables of the trapped electrons, $\tilde{f}_t(\theta,I)$, is not uniform in $\theta$. Indeed, trapping entails a shift in the angle, compared to a purely adiabatic evolution, that is essentially independent of the slowness of the dynamics. Consequently, the first Fourier component in $\theta$ of $\tilde{f}_t(\theta,I)$ is non zero, and depends very little on $\gamma$. Then, if the initial distribution in action was a Dirac distribution, $\si$ would oscillate in time, as shown in Ref.~\cite{action}. Hence, it would strongly differ from the adiabatic result plotted in Fig.~\ref{f1}. As is well-known, these oscillations in $\si$ translate into oscillations in the wave amplitude during the nonlinear stage of the cold beam-plasma instability~\cite{oneil}. Now, a smooth distribution, like the Maxwellian distribution used in Fig.~\ref{f1}, may be viewed as a (continuous) sum of weighted Dirac distributions, each providing its own oscillations in $\si$. When the distribution is smooth enough and the dynamics is sufficiently slow, eventually, all these dephased oscillations nearly average out to zero so that the contribution to $\si$ from the trapped electrons becomes negligible. Hence, although $\tilde{f}_t(\theta,I)$ is not homogeneous in $\theta$, its integral over a small interval in $I$ may be considered, after a long enough time, independent of $\theta$. The results of Fig.~\ref{f1} show that the small interval in $I$ is of the order of $m\vert \gamma\vert /k^2$, and that the long enough time is such that $\int \omega_B dt \agt 5$. Then, the condition on the smoothness of the distribution function and slowness of the dynamics translates into $\gamma/k\Delta v \ll 1$, where $\Delta v$ is the typical range in velocity over which the unperturbed distribution function varies significantly. For an initially Maxwellian plasma, the latter condition reads, $\gamma/kv_{th} \ll 1$ and, indeed, $\gamma/kv_{th}$ naturally appears as the relevant normalized growth rate when making use of dimensionless variables~\cite{benisti07}. Moreover, we also find that the adiabatic results eventually match the numerical ones only when $\gamma/kv_{th} \alt 0.1$, which supports the previous discussion. 
\begin{figure}[!h]
\centerline{\includegraphics[width=10cm]{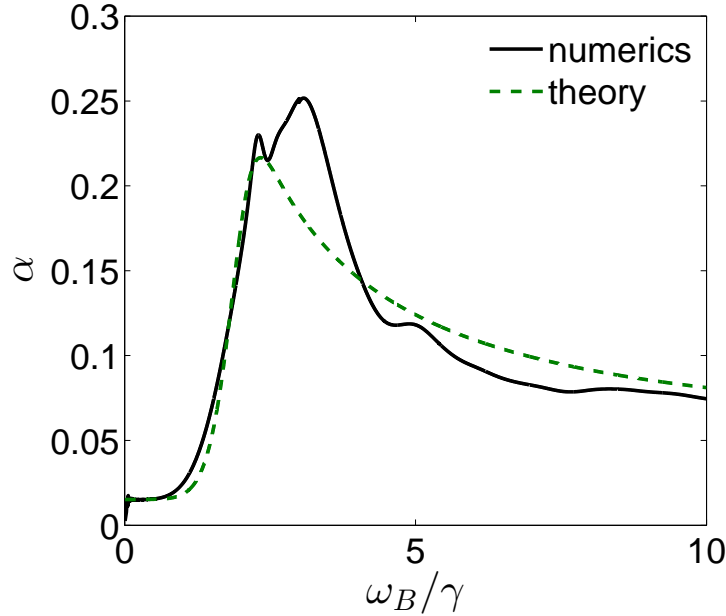}}
\caption{\label{f2}(color online)Values of $\alpha$ derived from numerical simulations (black solid line) and predicted theoretically (green dashed line) for the same parameters as in Fig.~\ref{f1}(b).}
\end{figure}

Now, let $\alpha$ be the term proportional to $E_1^{-2}\partial_t \omega$ in the envelope equation. According to the theoretical predictions of Paragraphs~\ref{new1}~and~\ref{new2}, $\alpha$ should be a constant in the linear regime, while, in the near-adiabatic one, to this constant should be added a term proportional $1/\sqrt{E_1}$. The latter term should be dominant at the transition from the linear to the near-adiabatic regime, since the transition occurs for small values of $E_1$. Now, about the transition between the two regimes, the contribution to $\si$ due to the nonzero value of $\partial_tÊ\omega$ is so small that, just from the results plotted in Fig.~\ref{f1}(b),  one cannot tell whether our theoretical prediction, regarding the sudden change in the way $\alpha$ should scale with $E_1$, is correct. In order to conclude, one needs a finer diagnostic. Then, let $S_{num}$ be the values of $-ne \langle \sin(\psi) \rangle/kE_1$ obtained numerically,  and let $S_0$ those inferred from Eq.~(\ref{n14})~\textit{without accounting for the term  proportional to}~$\partial_t \omega$. Since the accuracy of Eq.~(\ref{n14}) is excellent, $(S_{num}-S_0)/E_1\partial_t \omega$ provides a numerical estimate for $\alpha$ which, as illustrated in Fig.~\ref{f2}, is in good agreeement with theory.

The results plotted in Fig.~\ref{f2} are actually quite fascinating. They confirm the fact that, during  the sudden transition from the perturbative to the near-adiabatic regime, a new term does pop up in the wave equation, while this term was identically zero in the linear regime. To some extent, there is an analogy with phase transitions, where one defines an order parameter that is identically zero in one (disordered) phase, and suddenly assumes a finite value as the phase changes. Here, we can define a term in the wave equation that suddenly assumes a finite value when trapping becomes effective,  while it has absolutely no linear counterpart (unlike the term accounting for collisionless dissipation) and is, therefore, identically zero in the linear regime. This is as though the state of the EPW changed from a linear wave to a wave with trapped electrons. 

\section{General Theory}
\label{IV}
In this Section, we explain how the connection between the perturbative and near-adiabatic results, performed in Paragraph~\ref{new2}, generalizes to a non uniform wave. This leads to the wave equation~Eq.~(\ref{63})~in 1-D, and Eq.~(\ref{res})~in 3-D, which is our final result that we will discuss physically. 
\subsection{One-dimensional geometry}
\label{IV.1}
As shown in Section~\ref{new}, a very accurate envelope equation may be obtained by abruptly connecting the perturbative results with the near-adiabatic ones, as though the state of the wave abruptly changed. Furthermore, making use of first order results, which are close to the linear ones, proved to also yield quite an accurate wave equation. Here, we simply take advantage of these conclusions to derive a 1-D envelope equation. We look for an accurate equation, which, moreover, should be easy to implement in a code. Hence, the proposed equation simply amounts to connecting the linear results with the near-adiabatic ones. To do so we have to restrict to the situation when the wave packet is essentially bell shaped. If it composed of several well separated pulses, the situation is more complex because one has to account for the interactions between the various pulses, as recently investigated experimentally, numerically and theoretically in Ref.~\cite{pre_rousseaux}. Then, let $Y \equiv Y_1$ defined by Eq.~(\ref{n13}), and let $\Xi(x,t) \equiv \int_0^t \omega_B[\xi_\phi(t'),t'] dt'$, where $d\xi_\phi/dt=v_\phi$ and $\xi_\phi(t)=x$. We propose the following equation to describe the 1-D nonlinear propagation of an EPW in an inhomogeneous and non-stationary plasma,
\begin{eqnarray}
\nonumber
\varepsilon_0 E_1E_d/2-Y(\Xi)\int \mH_t \partial_x f_tdI =&-& Y(\Xi)\left[\partial_t\mL_{\omega}\vert_{\mA_s}-\partial_x\mL_k\vert_{\mA_s} \right] \\
\label{63}
&+&[1-Y(\Xi)]\left[\partial_t \partial_\omega \Lambda_{lin}-\partial_x\partial_k \Lambda_{lin}+2\nu_L \partial_\omega\Lambda_{lin} \right],
\end{eqnarray}
where we recall that $\Lambda_{lin} \equiv \varepsilon_0 \chi_{lin}E_1^2/4$, and $\mL_k \equiv \partial_k \mL$, $\mL_\omega \equiv \partial_\omega \mL$. 

The accuracy of Eq.~(\ref{63})~has been successfully tested against results from Vlasov simulations in Ref.~\cite{benisti09}~when the plasma was homogeneous. However, when the density is not uniform, an additional complexity arises as regards the definition of the electron distribution function. Indeed, $\chi_{lin}$ is defined in terms of the distribution function, $f_H$, while $\mL$ is expressed in terms of the ditribution functions $f_u$ and $f_t$, respectively, for the trapped and untrapped electrons. Hence, these distribution functions need to be connected, which is done the following way. First note that, in Eq.~(\ref{21}) ruling the evolution of $f_H$, one accounts for any slowly varying force. Since, in Ref.~\cite{benisti15}, only the linear limit was considered, this slowly varying force did not account for the term $\partial_x \mH_u$  (which, in the limit $E_1 \rightarrow 0$, is proportional to $E_1^2$ and corresponds to the usual ponderomotive force,  as shown  in Ref.~\cite{dodin13b}). However, since the derivation of Eq.~(\ref{res1})~actually consists in a linearization about the oscillation center, then, in 1-D, and neglecting the effect of collisions, one may just use the linear results of Paragraph~\ref{III.1} with $f_H(v)=mf_u(P/m)$. Now, the linear part of Eq.~(\ref{63}) is not expected to change much whether one uses $f_H(v)$ or $mf_u(P/m)$ since $\int \partial_x \mH_u dt$ should remain negligible when $\Xi \alt 5$. As regards the trapped distribution function, we already discussed in Paragraph~\ref{new1} that, when $\Xi \alt 5$, $f_t$ has to be defined so as to recover the linear dispersion relation in the small amplitude limit. If the plasma is homogeneous, this condition is automatically satisfied, while, if the plasma is inhomogeneous, one has to define $f_t$ so that the distribution function remains continuous through separatrix crossing. Only after $Y(\Xi)$ has become close enough to unity may one derive the evolution of $f_t$ from Eq.~(\ref{A6b}). Hence, there is some arbitrariness in the definition of the trapped distribution function, but whose impact on our description of wave propagation should be limited, as noted in Paragraph~\ref{new1}. Moreover, as we will now discuss it, the impact is further reduced by 3-D effects. 

\subsection{Three-dimensional geometry}
\label{IV.2}
\begin{figure}[!h]
\centerline{\includegraphics[width=10cm]{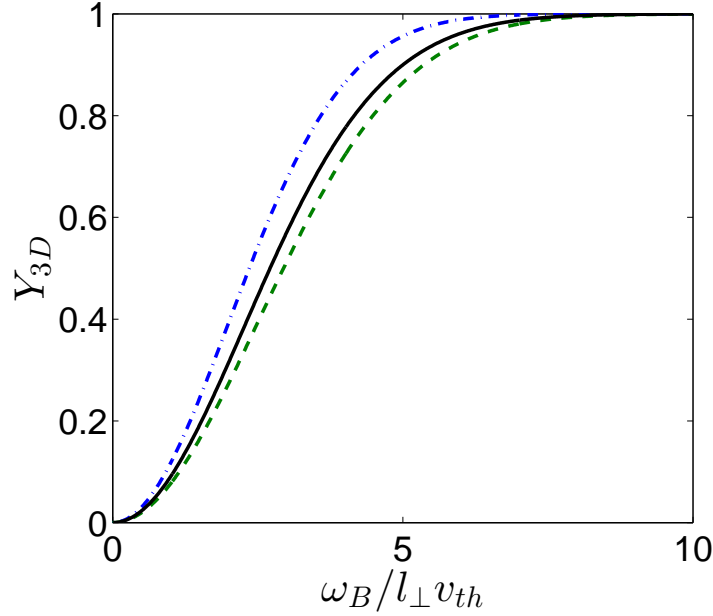}}
\caption{\label{f3}(color online)Values assumed by $Y_{3D}$ when $Y(\Xi_{3D})=\Upsilon(\Xi_{3D}/4-1)$ (blue dashed-dotted line), when $Y(\Xi_{3D})=\Upsilon(\Xi_{3D}/5-1)$ (green dashed line) and when $Y$ is defined by Eq.~(\ref{n13}) (black solid line). }
\end{figure}

Let us now generalize Eq.~(\ref{63}) to a three dimensional geometry. To do so, we have to assume that the EPW is localized in one given space location. In particular, we have to exclude the situation when the EPW results from the stimulated Raman scattering of a spatially smoother laser. Indeed, as discussed in recent papers~\cite{yin,prl_rousseaux}, such a situation would lead to complex couplings which are completely outside the scope of this paper. Then,  let us introduce at any point $M$,
\begin{equation}
\label{n20}
\Xi_{3D}(M) \equiv \int \omega_B[x_k(t'),y_k(t'),z_k(t')] dt',
\end{equation}
where $x_k$ is along the local direction of the wave number, and $dx_k/dt=v_\phi$, while $y_k$ and $z_k$ are along directions perpendicular to $\mathbf{k}$, and evolve in time according the dynamics transverse to $\mathbf{k}$. Moreover, $x_k(t)$, $y_k(t)$ and $z_k(t)$ are the local coordinates of the considered point, $M$. Note that, usually, the transverse dynamics is so weak that one may assume that the time evolution of $y_k$ and $z_k$ is just ballistic. Moreover, let us introduce
\begin{equation}
\label{67}
Y_{3D}Ê\equiv \int Y(\Xi_{3D}) f_{\bot}(v_{y_k},v_{z_k}) dv_{y_k}dv_{z_k},
\end{equation}
where $f_{\bot}(v_{y_k},v_{z_k})$ is the transverse electron distribution function, normalized to unity. In most cases, it is very close to the unperturbed one. Then, our proposed 3-D nonlinear envelope equation is just,
\begin{eqnarray}
\nonumber
\varepsilon_0 E_1E_d/2-Y_{3D}\int f_t \partial_{\mathbf{x}}.\frac{\mathbf{k}\mH_t}{k} dI =&-&Y_{3D}\left[\partial_t\mL_{\omega}\vert_{\mA_s}-\partial_{\mx}.\mL_{\mk}\vert_{\mA_s} \right] \\
\label{res}
&+&[1-Y_{3D}]\left[\partial_t \partial_\omega \Lambda_{lin}-\partial_{\mx}.\partial_{\mk} \Lambda_{lin}+2\nu_L \partial_\omega\Lambda_{lin} \right].
\end{eqnarray}
Now, the EPW is localized in a given space region, that the electrons cross due to their nonzero transverse velocity. Then, most often, the variations in $\Xi_{3D}$ are mainly due to this transverse motion. Now, clearly, as the electrons experience a growing wave, $Y_{3D}$ varies in a much smoother fashion than $Y$, leading to a much smoother transition in the way the EPW propagates. Moreover, it is also quite clear that the values assumed by $Y_{3D}$ weakly depend on the particular choice made for $Y$, provided that $Y$ fulfills the conditions derived in Section~\ref{new}, i.e., $Y\approx 0$ when $\Xi_{3D} <4$ and $Y \approx 1$ when $\Xi_{3D}>5$. Hence, one should obtain an upper bound for $Y_{3D}$ by choosing,  $Y(\Xi_{3D})=\Upsilon(\Xi_{3D}/4-1)$, $\Upsilon$ being the Heaviside function, and a lower bound by choosing, $Y(\Xi_{3D})=\Upsilon(\Xi_{3D}/5-1)$. Fig.~\ref{f3}~plots the values assumed by $Y_{3D}$ for each of the latter choices, and when $Y$ is given by Eq.~(\ref{n13}), in the situation when the change in the EPW electric field, as experienced by the electrons, is mainly due to its transverse gradient. Then, if we denote by $l_\bot$ the typical transverse scale length of variation of the EPW electric field, as long as the electrons experience a growing field, $\Xi \approx 2\omega_Bl_\bot/v_\bot$. 
The results plotted in Fig.~\ref{f3}~clearly show that our envelope equation, and therefore our predictions regarding the EPW propagation, depend very little on our choice for $Y$. This mainly removes the arbitrariness pointed out in Section~\ref{new}, in particular as regards the derivation of the trapped particles distribution function, $f_t$, and the nonlinear Landau damping rate, $\nu_{NL}$. Note that, from \res, we come to the remarkably simple formula $\nu_{NL}=\eta_{lin}Ê\nu_L$,  where $\eta_{lin}$ is the fraction of electrons that respond linearly to the wave. As already noted in Section~\ref{new}, this is just an effective damping rate, the complete description of nonlinear Landau damping requiring a much more complicated formulation than that provided by Eq.~(\ref{res}). Nevertheless, we know that the collisionless damping rate is more than $(1-Y_{3D})\nu_L$, when~$Y_{3D}$ is given by the blue dashed-dotted curve in Fig.~\ref{f3},~and~less than $(1-Y_{3D})\nu_L$ when $Y_{3D}$ assumes the values plotted by the green dashed line in Fig.~\ref{f3}. This yields quite a simple and accurate way to bound $\nu_{NL}$ from above and below. 

Now, still when the transverse variations of the EPW electric field are dominant, the linear electrons are the fastest ones, those which cross the domain enclosing the EPW within a time such that $\int \omega_B dt \alt 5$, and only such electrons significantly contribute to $\nu_{NL}$. However, if the transverse velocity of some electrons is so large that $E_1^{-1}\mathbf{v_\bot}.\mathbf{\nabla_\bot}E_1\agt\omega_{pe}$, these electrons do not experience a slowly-varying wave and do not respond adiabatically to it, unlike what has been assumed in this paper. Addressing non adiabatic electrons would require a completely different approach, more in the spirit of the recent paper~Ref.~\cite{action}, and is left for future work. Moreover, some electrons may be so fast that, for them, $\Xi_{3D} <5$ even when the wave is highly nonlinear and non sinusoidal, so that the result of Paragraph~\ref{III.1} do not apply. Hence, our theory is only valid when the contribution to the wave equation of such fast electrons is negligible. This is usually the case when $v_{th}/l_\bot k \lambda_D \ll \omega_{pe}$. Since $v_{th}/l_\bot$ is just the wave growth rate experienced by the fast electrons, the latter condition is no different from the condition, $\gamma/(k\lambda_D) \ll \omega_{pe}$, for a slowly-varying wave amplitude, which is one of our basic hypothesis. However,  as the wave grows, electrons have to be faster and faster to contribute to Landau damping so that, for large enough amplitudes, the nonzero value of $\nu_{NL}$ would be only due to the very fast, non adiabatic, electrons. Nevertheless, when the condition $v_{th}/l_\bot k \lambda_D \ll \omega_{pe}$ is fulfilled, the corresponding values found for $\nu_{NL}$ should be very small and, actually, smaller than the collisional damping rate. Then, our theory would need to be completed by the account for collisions. Addressing this issue in the nonlinear regime is outside the scope of this paper. However, if the fraction of trapped particles remains small, the expression Eq.~(\ref{22}) for the collisional damping rate should be valid. 

In spite of its limitations, we believe that~\res~yields the most general description of the EPW propagation, within the limits of geometrical optics, and for a smooth unperturbed distribution function. This equation encompasses such a rich physics, that it is certainly impossible to provide an exhaustive list of all effects it accounts for. Nevertheless, let us cite a few ones, which have already been discussed in previous publications. 

As shown in Ref.~\cite{dodinIII},~\res~would predict that, due to the space and time variations of $k$ and $\omega$, the EPW pulse would either split during its propagation, or its propagation would be unstable. The criterion for stable propagation, in the near-adiabtic regime, may be found in Ref.~\cite{dodinIII}, and is, approximately, $S<1/2$, where $S$ is the ratio between the contributions to $\partial_k \mL$ from trapped and untrapped electrons. 

\res~would also show that, in addition to Landau damping, the EPW pulse would shrink during its propagation, both longitudinally and transversally. As discussed in Ref.~\cite{dissipation}, this is due to the dissipation entailed by trapping. Moreover, solving~\res~together with the identity $\partial_t \mathbf{k}=-\mathbf{\nabla}Ê\omega$ and the nonlinear dispersion relation deduced from Eq.~(\ref{36})~or Eq.~(\ref{n1}), one would find that the EPW might self-focus during its propagation, as shown in Refs.~\cite{srs3D,yin08}

Furthermore, as discussed in Paragraph~\ref{new1}, if the wave phase velocity increases, the electron kinetic energy of the trapped electrons also increases, at the expense of the EPW,  whose amplitude should decrease as shown numerically in Ref.~\cite{schmit}. \res~does predict such a decrease in the EPW amplitude, as is clear from the sign of $\alpha$ plotted in Fig.~\ref{f2}. This may also be viewed as resulting from the conservation of the nonlinear wave action introduced in Paragraph~\ref{new4}~\cite{schmit}. Now, in a laser fusion device, an EPW driven by stimulated Raman scattering (SRS) would propagate towards the wall of the hohlraum, hence towards regions of higher density. As the EPW approaches the quarter critical density, its phase velocity increases, so that, eventually, the wave amplitude decreases and the EPW releases the electrons it has trapped. As discussed in Paragraph~\ref{III.2.2}, the distribution function of these detrapped electrons is symmetric with respect to the local phase velocity. Hence, when they are released, their  mean velocity is larger than their initial one, since $v_\phi$ has increased during the wave propagation. Consequently, a nonlinear plasma wave, driven by SRS in a fusion device, should create hot electrons, which is an issue for laser fusion. This issue may be addressed with the help of~\res, although an accurate description for the production of hot electrons would require to include more physics in the model, and, in particular, to account for the potential growth of sidebands, as further discussed in the conclusion.

\subsection{Discussion on the definition of the wave action and on the concept of  plasmon}
\label{new4}
Our final envelope equation,~\res, results from the connection of linear and adiabatic formulas and, consequently, it is expressed in terms of two different functions, $\mL$ and $\Lambda_{lin}$. It is certainly possible to find an \text{eff}ective Routhian density, $\mL^{\text{eff}}$, so that the right-hand side of~\res~would just be $\partial_t \left.\mL^{\text{eff}}_\omega\right\vert_{\mA_{\max}}-\partial_{\mx}.\left.\mL^{\text{eff}}_{\mk}\right\vert_{\mA_{\max}}$, where $\mL^{\text{eff}}_\omega\equiv \partial_\omega \mL^{\text{eff}}$ and $\mL^{\text{eff}}_{\mk}\equiv \partial_{\mk} \mL^{\text{eff}}$. Actually, $\mL^{\text{eff}}$ would just be the Routhian density one would derive from a nonlocal variational formalism, valid whatever the regime of wave-particle interaction. However, even if such a Routhian density could be found, $\mL^{\text{eff}}_\omega$ would not provide a good definition for the wave action density. This was already outlined in Paragraph~\ref{III.1}~in the linear regime, where we showed that $\partial_\omega \mL_{lin}$ remained constant while the wave was Landau damped, and it does make quite sense to define a constant action for a decaying wave. 

One way to alleviate this difficulty would be to remove, from the frequency derivative of the Routhian density, the terms accounting for dissipation.  In the linear regime, this would lead to $\partial_\omega \Lambda_{lin}$ as a definition for the action density, which is the usual one. Then, Eq.~(\ref{res1}) would just express the Landau and collisional damping of the wave action, for a freely propagating wave ($E_d=0$). Similarly, in Paragraph~\ref{new1}, we discussed the fact that, mathematically speaking, collisionless dissipation due to trapping naturally appeared in the wave equation because, for trapped electrons, $\mh$ did not depend on the current frequency. However, from the definition Eq.~(\ref{A9b})~of the dynamical action of trapped electrons, $\mh$ depends on all previous values of $\omega$. Then we introduce, $\partial \mL/\partial  \omega_{all}$, the derivative of $\mL$ with respect to all the frequencies that appear in its expression, the current one as well as all previous ones, and we define the action density, $\mI$, so that 
\begin{equation}
\label{a1}
\frac{\partialÊ\mI}{\partial t}=(1-Y)\frac{\partial^2Ê\Lambda_{lin}}{\partial t\partial \omega}-Y \frac{\partial \mL}{\partial t \partial \omega_{all}}. 
\end{equation}
Now, from Eq.~(\ref{E13e}), and using the same kind of calculations as in Paragrah~\ref{new1}, one easily finds that $-\partial^2 \mL/\partial E_1 \partial \omega_{all}$ converges towards $\partial^2 \Lambda_{lin}/\partial E_1 \partial \omega$ when $E_1 \rightarrow 0$ [provided that $F'_t(v_\phi)=F'_u(v_\phi)$]. Hence, when $k$ and $\omega$ are constant, the action density defined by~Eq.~(\ref{a1}) would continuously vary from the linear to the adiabatic regimes. By contrast, if $k$ and $\omega$ may vary, the action density would abruptly change when shifting from the linear to the adiabatic regime, due to terms similar to $\alpha$, plotted in Fig.~\ref{f2}. Clearly this is because, in the adiabatic regime, the variation of the kinetic energy of trapped electrons enters into the definition of $\mI$, as is obvious from the expression~Eq.~(\ref{31})~for $\mH_t$. The action density being homogeneous to an energy density per frequency, it is often related to the density of plasmons~\cite{pines}, i.e., the density of quanta for the EPW. Therefore, interestingly enough, in 1-D we would find an abrupt change in the density of plasmons when the EPW propagation changes from linear to nearly adiabatic. Once again, this would be reminiscent of a phase transition. 

Let us now discuss the relevance of the concept of plasmon. Using Eq.~(\ref{a1}) for the action density, we would find that the total action would vary due to Landau damping, due to the dissipation induced by trapping, and due to the inhomogeneous loading of trapped electrons, leading to the term $\int \partial_x\mH_t f_t dI$. As already discussed in Paragraph~\ref{new1}, Landau damping cannot be recovered in the adiabatic limit, hence, it is entailed by the non adiabaticity of the electron motion and, therefore, by the non conservation of their dynamical action, $I$. Similarly, as pointed out in Paragraph~\ref{III.2.2}, the dissipation due to trapping results from a geometric change in the dynamical action, $I$, by the width of the resonance. The third and last term that makes the wave action vary is inhomogeneity, and it also entails a change in the dynamical action of the untrapped electrons. Hence, we conclude that the wave action varies whenever the electron dynamical action is not conserved. This makes sense, because a plasma wave is nothing but density fluctuations. Therefore, if one cannot find an adiabatic invariant for the electron motion, such an invariant should not exist as regards the EPW propagation either. Moreover, we also come to the conclusion that the number of plasmons is not a true invariant, its invariance is broken by non adiabaticity. Therefore, the concept of plasmon seems questionable, it is clearly not as robust as the concept of photon, for example. 

\section{conclusion}
\label{V}
In this paper, we derived, for the first time, an envelope equation that accounts for collisionless dissipation and for the plasma inhomogeneity and nonstationarity. Hence, we believe that, within the limits of geometric optics, and for a smooth unperturbed distribution function,~\res~describes the propagation of an EPW in the most general way. This equation has been derived thanks to a careful investigation of wave-particle interaction at a microscopic level, that was coupled with the effects induced by the variations of the wave amplitude, and of the plasma, over long space and time scales. To do so, we resorted to the linear results derived in Ref.~\cite{benisti15}~and to the variational method first introduced in Ref.~\cite{dodin12} and generalized in this article in order to allow for trapping and detrapping. This led us to introduce a Lagrangian density that was nonlocal in the wave number, frequency and amplitude, so as to account for the collisionless dissipation induced by trapping. 

Now, there are many known limitations to describing wave propagation using a first order differential equation such as~\res. To cite a few examples, the EPW may by subjected to secondary instabilities leading to the growth of sidebands~\cite{kruer,brunner, dodin13}, so that, except in some special instances~\cite{keen1,keen2}, the total electrostatic field, $\mathbf{E}(\mathbf{x},t,\psi)$, is no longer a slowly-varying function of $\mathbf{x}$ and $t$ (at fixed $\psi$). Moreover, due to the amplitude dependence of the wave frequency, the wave front may bend so as to make  the EPW self-focus~\cite{yin08,srs3D}, which, again, renders the assumption of a slowly-varying amplitude no longer valid. In the situation, observed both numerically and experimentally~\cite{yin08,prl_rousseaux_09}, when wavefront bowing is followed by the growth of sidebands, the phase modulation due to the transverse variations of the wave amplitude is negligible compared to that due to the space dependence of the nonlinear frequency shift. Consequently,  in such a situation, diffraction-like effects become inessential. Similarly, as discussed in Ref.~\cite{vgroup}~and in Paragraph~\ref{III.2.2}, due to trapping the EPW group velocity varies in a nonlinear and nonlocal fashion, and these variations may dominate the effect of group velocity dispersion. Hence, known limitations of a first order wave equation in the linear regime may become mostly irrelevant as regards the nonlinear EPW propagation, which gives more credit to~\res~than could be inferred from previous studies on wave propagation. However this equation is, clearly, only one first step towards a more complete description of the nonlinear propagation of an EPW. Nevertheless, this this first step cannot be avoided, and it happens to be quite relevant to address further nonlinear issues. Indeed, in Ref.~\cite{friou}, a 1-D envelope code~\cite{brama}, solving the restriction of~\res~to a homogeneous and stationary plasma,  was used to predict when sidebands could  reach significant amplitudes and stop the coherent growth of SRS. The predictions of this code were in very good agreement with those obtained with Vlasov simulations. The 3-D version of this code was used in Ref.~\cite{srs3D}~to predict when the EPW would self-focus and, again, saturate SRS, with predictions on Raman reflectivity comparing even better with experiment~\cite{montgomery}~than those from PIC codes~\cite{yin08}. 

Moreover~\res~is useful \textit{per se} in order to address practical issues, make comparisons with basic physics experiments or test theories. Indeed, as mentioned above, envelope equations proved to provide good estimates of Raman reflectivity, in a homogeneous plasma, in 3-D, and in the strongly nonlinear regime. However, to make predictions regarding the important issue of laser fusion, one needs to account for the plasma inhomogeneity. \res~is one important step in that direction. It now needs to be coupled with the propagation of the laser and backscattered waves, which will be the subject of a forthcoming paper. Furthermore, as discussed in Paragraph~\ref{IV.2},~\res~may be used to predict the production of hot electrons which also is a concern for fusion. 

As regards basic physics, we provided, in this article, a simple theoretical formula for the nonlinear Landau damping rate. It is simply proportional to the fraction of electrons that cross the EPW in a time less than, about, one bounce period. Hence, as the wave grows, the Landau damping rate should decrease smoothly, which is in qualitative agreement with the experimental results reported in Ref.~\cite{danielson}. However, in these experiments, one observes early oscillations in the wave amplitude before damping, which seems to indicate that the wave has grown in a sudden way, and not adiabatically as assumed in this paper. Hence, we cannot get into quantitative comparisons here, which would, moreover, require  the knowledge of the longitudinal and transverse profile of the EPW. However, experimental testing of our theory would be welcome, even at a qualitative level one could check whether, as predicted here, the Landau damping rate would fall more rapidly, as the wave amplitude increases, when the transverse extent of the wave is larger, and the geometry closer to 1-D.

Solving an equation as complex as~\res, even numerically, is a work in itself. This is clearly outside the scope of this paper, but will be the subject of a future article. However, solving~\res~would allow to test theoretical predictions regarding nonlinear waves. Namely, it would allow to study the type of nonlinear waves one could obtain, depending on the way they have been driven. For example, in Ref.~\cite{schamel}, it has been proven that cnoidal waves were stationary solutions of the Vlasov-Poisson system. Then, one could study whether~\res, together with Eq.~(\ref{36}), would predict that such waves may, indeed, be excited, and the way they would propagate.  
\begin{acknowledgments}
The author gratefully acknowledges numerous useful discussions with I.Y. Dodin, and the hospitality of the Princeton Plasma Physics Laboratory in July 2015. 
 \end{acknowledgments}

\appendix
\section{Action-angle variables}
\label{A}
\setcounter{equation}{0}
\newcounter{app}
\setcounter{app}{1}

In this Appendix we quickly recall, for the sake of definiteness, known results on action-angle variables. 

The Hamiltonian, $H$, ruling the electron dynamics in $(x,p)$ variables, where $p \equiv m\dot{x}-eA$, is~\cite{jackson},
\begin{equation}
\label{A1}
H=\frac{(p+eA)^2}{2m}+\Phi,
\end{equation}
where the potential $\Phi(x,t)$ reads, $\Phi(x,t) \equiv \Phi[x,t,\psi(x,t)]$. It is $2\pi$-periodic in $\psi$ (at fixed $x$ and $t$) and varies slowly with $x$ and $t$ (at fixed $\psi$), and the same holds for $A$. Let us now shift to $(\psi,p')$ variables, using the generating function $G \equiv \psi p'$, which yields $p=\partial G/\partial x=kp'$. In $(\psi,v')$ variables, the Hamiltonian is $\mH_e=H+\partial G/\partial t = H-\omega p'$, which reads
\begin{equation}
\label{A2}
\mH_e=\frac{(kp'+eA')^2}{2m}+\Phi'-\omega p',
\end{equation}
where $A'(\psi,t) \equiv A[x(\psi,t),t]$ and $\Phi '(\psi,t) \equiv \Phi[x(\psi,t),t]$.

When the wave amplitude varies in space and time, the notion of trapped or untrapped electrons becomes ambiguous. Then, as regards the definition of the action and angle variables, we adopt the following convention. An electron is considered as untrapped if, for the frozen dynamics (obtained by replacing the time-dependent Hamiltonian by a constant one, corresponding to the value assumed by $\mH_e$ at the considered time), the electron orbit may span an interval in $\psi$ larger than $2\pi$, about the considered space location. For these ``untrapped'' electrons, we define the action, $I$, by,
\begin{equation} 
\label{A3}
I(\mH_e,\psi,t) \equiv \frac{1}{2\pi} \int_{\psi-\pi}^{\psi+\pi} p'(\mH_e,\psi',t)d\psi',
\end{equation}
where $p'(\mH_e,\psi',t)$ solves Eq.~(\ref{A2}), and where the integral is calculated over a frozen orbit, i.e., for a given, fixed, value of $\mH_e$. Provided that the relation between $I$ and $\mH_e$ is invertible, $\mH_e$ reads, $\mH_eÊ\equiv \mH_e(I,\psi,t)$. Note that the $\psi$-dependence of $I$ is only due to the, slow, non-periodic space-dependence of $\Phi$ and $A$. Consequently the action, $I$, varies slowly with $\psi$, and the same is true for $\mH_e$. Let the angle, $\theta$, be canonically conjugated to $I$. Then, the generating function of the change of variables $(\psi,p') \rightarrow (\theta,I)$ is $g(\psi,I,t) \equiv \int p'(I,\psi,t) d\psi$, and $\theta = \partial g/\partial I$. Note that a variation by $2\pi$ in $\psi$ amounts to a variation by $2\pi$ in $\theta$, and a variation by $2\pi I$ in $g$. In action-angle variables, the Hamiltonian is 
\begin{equation}
H'_e(I,\theta,t)=H_e(I,\theta,t)+\partial g/\partial t\vert_I,
\end{equation}
where $H_e(I,\theta,t) \equiv \mH_e[\psi(I,\theta),I,t]$. Then, 
\begin{eqnarray}
\nonumber
\frac{dI}{dt}&=&-\left.\frac{\partial H'_e}{\partial \theta}\right\vert_I,\\
\label{A4}
&=& \left.-\frac{\partial \mH_e}{\partial \psi}\frac{\partial \psi}{\partial \theta}\right\vert_I-\left.\frac{\partial^2g}{\partial t\partial \theta}\right\vert_I.
\end{eqnarray}
Let us now average Eq.~(\ref{A4}) over a $2\pi$-interval in $\theta$, and for a fixed value of $I$, in order to derive what we denote by $\langle dI/dt \rangle_\theta$. Since $\partial \mH_e/\partial \psi$ varies slowly with $\psi$, this factor may be considered as a constant during the averaging procedure. Moreover, since a variation by $2\pi$ in $\theta$ entails a variation by $2\pi$ in $\psi$, $\langle \partial \psi/\partial \theta\rangle=1$. Hence, at lowest order in the space-dependence of the potentials, the first term just yields $-\partial \mH_e/\partial \psi$. As for the second term, it just averages out to 0 because $\langle \partial g/\partial \theta\rangle_\theta=2\pi I$, and the time derivative is calculated at fixed $I$. Hence, Eq.~(\ref{A4}) yields,
\begin{equation}
\label{A5}
\langle dI/dt \rangle_\theta \approx -\partial \mH_e/\partial \psi.
\end{equation}
Note also that, due to the periodicity of $g$, $\langle d\theta/dt \rangle_\theta =\partial_I \mH_e$.  \\

For trapped electrons, the action is defined as 
\begin{equation}
\label{A6}
I_{\psi_O}(\mH_e,t) \equiv \frac{1}{2\pi} \oint p'(\mH_e,\psi,t) d\psi,
\end{equation}
where the integral is taken over a frozen orbit. This orbit encircles the $O$-point of the resonance (i.e., the trapped island), whose abscissa, $\psi_O$, is, clearly, a constant. Hence, for trapped orbits, there is one set of action-angle variables per resonance, and each set is labelled by $\psi_O$. Moreover, as is clear from Eq.~(\ref{A6}), for one given $\psi_O$, the action is independent of $\psi$. Consequently, so is $\mH_e$, and Eq.~(\ref{A5})  just yields $\langle dI_{\psi_O}/dt \rangle_\theta=0$. Actually, since the motion of trapped electrons is genuinely periodic, $I_{\psi_O}$ is an adiabatic invariant. $\vert I_{\psi_O}(t)-I_{\psi_O}(0) \vert \alt \varepsilon$ provided that $t \alt 1/\varepsilon$, where $\varepsilon \sim \vert (\omega \mH_e)^{-1} \partial_t \mH_e\vert$~(see, for example, Ref.~\cite{arnold}). Consequently, the distribution function of trapped electrons, $f_t$, is nearly conserved, $f_t[\psi_O(t),I_{\psi_O}]\approx f_t[\psi_O(0),I_{\psi_O}]$. Now, remember that we only aim at deriving envelope equations which are averaged over one wavelength. Therefore, at the order where we work, $f_t(\psi_O,I_{\psi_O}) \approx f_t(\psi',I_{\psi_O})$, provided that $\vert \psi_O-\psi'Ê\vert<2\pi$. This allows to remove the tag associated to each resonance, and to consider the distribution of trapped electrons as a continuous function of $\psi$. Then, denoting now the action of a trapped orbit simply by $I_t$, one finds,
\begin{equation}
\label{A6b}
f_t[(\psi(t),I_t]Ê\approx f_t[(\psi(0),I_t].Ê
\end{equation}

Now, as regards the main result, Eq.~(\ref{A5}), there are two ways of interpreting it. The first point of view, adopted in Refs.~\cite{dodin12,dodin13b}, consists of considering the averaging over $\theta$ as a time averaging over the period, $T$, of a frozen orbit. Then, Eq.~(\ref{A5}) may only be accurate if $I$ remains nearly constant for a time interval larger than $T$. However, close to the frozen separatrix, $T \rightarrow \infty$. Hence, only for electrons whose orbit remains far away from the separatrix, may Eq.~(\ref{A5}) provide a correct estimate of the time-averaged variation of the action, $I$. Moreover, since $I$ is assumed to have changed very little during one period, Eq.~(\ref{A5}) may be approximate by
\begin{equation}
\label{A5b}
dI/dt \approx -\partial \mH_e/\partial \psi.
\end{equation}

The second point of view consists in considering the averaging over $\theta$ as a space averaging over one frozen orbit. Then, because the action remains nearly constant over one wavelength, Eq.~(\ref{A5}) may, again, be approximated by Eq.~(\ref{A5b}). Now, \textit{an averaging over $\theta$ may be viewed as a space averaging only if the distribution in $\theta$ is nearly uniform}. For one given $I$, this is not true close to the separatrix, or after the orbit has been trapped~\cite{action}. However, from the results of Paragraph~\ref{new3} we know that, for times such that $\int \omega_B dt >5$, and provided that the action has remained nearly constant for all $t$, phase mixing has become so effective that, on the average over a small interval, $\delta I$, the distribution in $\theta$ may be considered as uniform. As discussed in Paragraph~\ref{new3}, $\delta I \sim m\vert \gamma \vert/k^2$, where $\gamma$ is the rate of variation of the wave amplitude (in the reference frame moving at the phase velocity). For the slowly varyings waves considered in this paper, $\delta I$ is a very small interval so that, if  the distribution in action is smooth enough, the distribution in angle may, effectively, be considered as uniform. This is confirmed by the results plotted in Fig.~\ref{f1}, so that, in practice, one may consider that Eq.~(\ref{A5}) yields the space averaged evolution of each $I$. Now, demanding that the action remains nearly constant for all times seems to be a more stringent condition than requiring that $I$ remains nearly constant during one orbit period, $T$, as needed to interpret Eq.~(\ref{A5}) as a time averring. Nevertheless, if $I$ has changed a lot due to inhomogeneity (that affects all values of $I$ nearly the same way), then, from Eq.~(\ref{30}) on the evolution of the distribution function in $(X,kI)$ coordinates, nothing guarantees that the distribution in $X$ would remain slowly varying, which would break our main hypothesis. Actually, Eq.~(\ref{A5b}) is only useful for a finite time, as long as $I$ has not changed too much. Indeed, since the passing orbits are not periodic, nothing guarantees that the averaged evolution would be representative of the true one~\cite{arnold}. 
\\

Before ending this Appendix, let us slightly reformulate our definitions and results. In order to keep the same notations as in Refs.~\cite{dodin12,dodin13b}, we introduce,
\begin{equation}
\label{A7}
\mh \equiv k^2(p'-mu/k)^2/2m+\Phi',
\end{equation}
where $u \equiv v_\phi-eA'/m$ (and $v_\phi \equiv \omega/k$). Then, $\mH_e$ reads,
\begin{equation}
\label{A8}
\mH_e = \mh-mu^2/2+e^2A^2/2m,
\end{equation}
so that $\mh$ may be used as an equivalent Hamiltonian when $A'=0$ and $v_\phi$ is uniform. From Eq.~(\ref{A7}), the action for the untrapped electrons may be written as
\begin{equation}
\label{A9}
I_u=\frac{mu}{k}+\frac{\eta\sqrt{2m}}{2\pi k}Ê\int_{\psi-\pi}^{\psi+\pi} \sqrt{\mh-\Phi} d\psi',
\end{equation}
where $\eta$ is the sign of $(kp'/m-v_\phi)$. The integral in Eq.~(\ref{A9}) is to be calculated for a given value of $\mH_e$. However, at the order where we work, the integral may also be calculated for a fixed value of $\mh$. 

Moreover, we want to define the action so that it remains continuous through separatrix crossing. Then, following Ref.~\cite{benisti07}, we  define the action of a trapped orbit by
\begin{equation}
\label{A9b}
I_t \equiv \frac{mu_0(I_t)}{k}+\eta_0(I_t)\frac{\sqrt{2m}}{4\pi k}Ê\oint  \sqrt{\mh-\Phi} d\psi',
\end{equation}
where $u_0(I_t)$ are $\eta_0(I_t)$ are the values assumed by $u$ and $\eta$ when the orbit has been trapped
. Note that, with this definition for $I_t$, the same trapped orbit in the $(x,v)$ space is split into two distinct obits in the $(I,\theta)$ space (see Ref.~\cite{DNC}). However, when phase mixing is effective, in practice, one may consider that the distribution is the same, and uniform, on both orbits. Only then may a space averaging over one wavelength be identified with a $2\pi$ averaging over a uniform distribution in $\theta$. Note also, from Eqs.~(\ref{A9}) and (\ref{A9b}),  that an orbit becomes trapped when
\begin{equation}
\label{u}
\vert I_u- mu/k \vert < A_s/4\pi \text{ and }\frac{d}{dt}(\mA_s/4\pi+\eta mu/k)>0,
\end{equation}
where $\mA_s$ is the area of the separatrix. Moreover, an orbit remains trapped as long as
\begin{equation}
\label{t}
\vert I_t- mu_0/k \vert < A_s/4\pi.
\end{equation}
Note that, if $\vert I_u- mu/k \vert < A_s/4\pi$ and $ d(\mA_s/4\pi+\eta mu/k)/dt\leq0$, the orbit remains untrapped, but its action varies by $\pm \mA_s/4\pi$~\cite{ratchet}. 

Finally, in order to make contact with the results of Refs.~\cite{dodin12,dodin13b}, let us introduce, $P\equiv kI$. Then, from Eq.~(\ref{A5b}) one easily finds, 
\begin{equation}
\label{A10a}
\langle dP/dt \rangle=-\partial_x \mH_u,
\end{equation}
with
\begin{equation}
\label{A10}
\mH_u=\mh+Pv_\phi-mv_\phi^2/2-eAv_\phi.
\end{equation}

\section{Landau damping from a variational formalism.}
\label{B}
\setcounter{app}{2}
As discussed in Paragraph~\ref{III.1}, we only give here the derivation of Landau damping in a simple situation, when the plasma and the wave amplitude are assumed to be homogeneous and when $k$ and $\omega$ are constant. Moreover, in the linear regime, we may assume that the electric field, $E$, is sinusoidal and that it derives from a scalar potential. Hence, $E$ reads, $E \equiv (E_1/2) e^{i\psi}+c.c.$, and the Hamiltonian for wave-particle interaction is
\begin{equation}      
\label{B1}
H_e=\frac{p^2}{2m}-\left(\frac{\Phi_1}{2}e^{i\psi}+c.c.\right),
\end{equation}
where $E_1$ is related to $\Phi_1$ by, $-eE_1=ik\Phi_1$. 

Now, in the expression~Eq.~(\ref{23}) for the Lagrangian density, one may use any set of canonically conjugated variables, $(x',p')$, to express the Lagrangian, $L_e$, and, therefore, for the Hamiltonian $H_e$. Here, we choose to define $(x',p')$ as resulting from a perturbative expansion, in the potential amplitude, $\Phi_1$. Namely, we define
\begin{eqnarray}
\label{B2}
x'& \equiv &x+\partial_{v'} F,\\
\label{B3}
p&\equiv &p'+\partial_x F,
\end{eqnarray}
where the generating function, $F$, is expanded in powers of $\Phi_1$, $F\equiv \sum_n \Phi_1^n F_n$. Then, in the new variables, $(x',p')$, the new Hamiltonian is $H' \equiv H+\partial_t F$. Here, we try to look for a generating function, $F$, that would cancel out the potential part of $H$, so that $F$ would be the solution to,
\begin{equation}
\label{B4}
v'\partial_x F+\partial_t F +(\partial_x F)^2/2m=(\Phi_1/2)e^{i\psi}+c.c.,
\end{equation}
where $v'Ê\equiv p'/m$. At first order in the wave amplitude, Eq.~(\ref{B4}) reads,
\begin{equation}
\label{B5}
v'\partial_x F_1+\partial_t F_1=(\Phi_1/2)e^{i\psi}+c.c.
\end{equation}
Eq.~(\ref{B5}) is easily solved and yields
\begin{equation}
F_1=\frac{1}{2} \int_0^t \Phi_1(t') e^{i\psi[(x-v(t-t'),t']}dt'+c.c.
\end{equation}
Then,
\begin{equation}
\frac{\partial F_1}{\partial x}=\frac{-e}{2} \int_0^t E_1[(t') e^{i\psi[(x-v(t-t'),t']}dt'+c.c.,
\end{equation}
and, at first order in the perturbation analysis, the new Hamiltonian is
\begin{equation}
H'_1=\frac{p'^2}{2m}+\frac{1}{2m}\left( \frac{\partial F_1}{\partial x}\right)^2.
\end{equation}
It is the sum of rapidly varying terms, that may be ``killed'' at higher order in the perturbative expansion, and of a slowly varying term that is not perturbative and, therefore, needs to be kept. Then, in the variables $(x',p')$, the new Hamiltonian reads
\begin{equation}
\label{B6}
H'=\frac{p'^2}{2m}+\left\{\frac{e^2}{8m}\int_0^t \int_0^t E_1(t')E_1^*(t'')e^{i\left\{\psi[\xi(t'),t']-\psi[(\xi(t''),t'']\right\}}dt'dt''+c.c\right\}+O(E_1^4),
\end{equation}
where $\xi(\tau) \equiv x-v'(t-\tau)$, and $x$ is implicitly considered as a function of $x'$ and $p'$, $x\equiv x(x',p',t)$.  Here, we only look for a linear result and, consequently, we  henceforth neglect the $O(E_1^4)$ term in Eq.~(\ref{B6}). Since $H_1$ is quadratic in the wave amplitude, at first order, $v'$ is a constant of motion. Hence, the electron distribution function in $(x',v')$ is just what we denoted in Paragraph~\ref{III.1}, $f_H$. Moreover, since the change of coordinates $(x,p) \rightarrow (x',p')$ is canonical, $f_H(x',v',t) = f(x,v,t)$. Then, since the plasma is homogeneous, the Lagrangian density defined by Eq.~(\ref{23}) reads, 
\begin{equation}
\label{B7a}
L=\varepsilon_0 \frac{E^2}{2}+\left\{\int_{-\infty}^{+\infty} f_H(v',t)\left\{mv'/2-H'[x'(x,v,t),v',t]\right\}\frac{\partial x'}{\partial x}dv'+c.c.\right\}
\end{equation}
From Eq.~(\ref{B6}), at lowest order in the wave amplitude, $mv'/2-H'[x'(x,v,t),v',t] \equiv -\mH(x,v',t)$, where,
\begin{equation}
\label{B7}
\mH(x,v',t) = \frac{e^2}{8m}\int_0^t \int_0^t E_1(t')E_1^*(t'')e^{i\left\{\psi[\xi(t'),t']-\psi[(\xi(t''),t'']\right\}}dt'dt''+c.c.,
\end{equation}
with $\xi(\tau) = x-v'(t-\tau)$. Then, in the spirit of the multiple scale analysis,
\begin{equation}
\label{B8}
L=\varepsilon_0 \frac{E^2}{2}-\int_{-\infty}^{+\infty} f_H(v',t) \mH[x,v',t]\frac{\partial x'}{\partial x}dv',
\end{equation}
reads $L \equiv L(x,t,\psi)$, and is periodic with respect to $\psi$ (at constant $x$ and $t$). Hence, from Whitham theory~\cite{whitham}, one may apply Lagrange equations to $\langle L \rangle$, which is the averaged value of $L$ over $\psi$ (at constant $x$ and $t$), and which is straightforwardly found to be,
\begin{equation}
\label{B9}
\langle L \rangle = \varepsilon_0 \frac{\vert E_1\vert ^2}{4}-\int_{-\infty}^{+\infty} f_H(v',t) \mH[x,v',t]dv'.
\end{equation}

In order to derive the integral in the right-hand side of Eq.~(\ref{B9}) we use the following identity,
\begin{equation}
\label{B10}
\int_{-\infty}^{+\infty} f_H(v') \mH(x,v') dv'= \int_{-\infty}^{+\infty} [f_H(v')-f_H(v_\phi)]\mH(v')dv'+f_H(v_\phi)\int_{-\infty}^{+\infty} \mH(v')dv'.
\end{equation}
Now, 
\begin{eqnarray}
\nonumber
\int_{-\infty}^{+\infty} \mH(v')dv'&=&\frac{e^2}{8m}\int_0^t\int_0^t E_1(t')E_1^*(t'')\left[\int_{-\infty}^{+\infty} e^{i\left\{\psi[\xi(t'),t']-\psi[(\xi(t''),t'']\right\}}dv' \right]dt'dt''+c.c. \\
\nonumber
&=&\frac{e^2}{8m}\int_0^t\int_0^t E_1(t')E_1^*(t'') (2\pi/k) \delta (t'-t'')+c.c. \\
\label{B13}
&=&  \frac{\pi e^2 f_H(v_\phi)}{2m}\int_{0}^t\frac{\vert E_1 \vert^2(t')}{k}dt'.
\end{eqnarray}
Moreover, as shown in Ref.~\cite{benisti15}, at zero order in the variations of $E_1$, one may just estimate the first term in the right-hand side of Eq.~(\ref{B10}) by replacing, in the time integral Eq.~(\ref{B7}), $E_1(t')$ and  $E_1(t'')$ by $E_1[\xi(t),t]=E_1(x,t)$. Then, one easily finds,
\begin{eqnarray}
\nonumber
\int_{-\infty}^{+\infty} [f_H(v')-f_H(v_\phi)]\mH(v')dv'& \approx & \frac{e^2 \vert E_1 \vert^2(x,t)}{4m}P.P \left(\int\frac{f_H(v)-f_H(v_\phi)}{(kv-\omega)^2}dv\right), \\
\label{B14}
&=& -\varepsilon_0 \chi_{lin}(k,\omega) \vert  E_1Ê\vert^2/4.
\end{eqnarray}
Using Eqs.~(\ref{B9}),~(\ref{B10}),~(\ref{B13})~and~(\ref{B14}), one finds the following averaged Lagrangian density, $\mL_{lin}Ê\equiv \lL$, such that
\begin{equation}
\label{B15}
\mL_{lin} \approx  \varepsilon_0 (1+\chi_{lin}) \frac{\vert E_1 \vert^2}{4}-\frac{\pi e^2 f_H(v_\phi)}{2m}\int_{0}^t\frac{\vert E_1 \vert^2(t')}{k}dt',
\end{equation}
which is, clearly, non local in the wave amplitude, and is also to be considered as non local in $k$, although  $k$ is assumed to remain constant. Hence, to proceed, we introduce the amplitude $A_0 \equiv \vert E_1 \vert ^2/k$, so that the Lagrangian density now reads, 
\begin{equation}
\label{B16}
\mL_{lin} \approx  \varepsilon_0 (1+\chi_{lin}) k\frac{A_0}{4}-\frac{\pi e^2 f_H(v_\phi)}{2m}\int_{0}^tA_0(t')dt',
\end{equation}
and is local in $k$ (although still non local in $A_0$). Then, Lagrange equation  $\partial_{t \omega}Ê\mL_{lin}=0$ reads,
\begin{equation}
\partial_\omega \chi_{lin} (\partial_t \vert E_1 \vert^2+2\nu_L \vert E_1 \vert^2)=0,
\end{equation}
 where $\nu_L$ is the Landau damping rate, defined by Eq.~(\ref{21}). Hence, we do find, using a variational formalism, that the EPW is Landau damped.

\section{Averaged variational principle}
\label{C}
\setcounter{app}{3}

In a spirit close to Whitham's~\cite{whitham}, let us now discuss the validity of Lagrange equations written for the space averaged Lagrangian density, over one wavelength, which we denote by $\langle L \rangle$. 

From Eq.~(\ref{23}), the Lagangian density is,
\begin{equation}
\label{C1}
L(x_0,t)=\varepsilon_0 E^2/2+\int \delta(x-x_0) L_e(x,v,t) f(x,v,t)dxdv,
\end{equation}
 where $L_e$ is given by Eq.~(\ref{23b}). Now, in order to use Lagrange equations in an unambiguous fashion, one should use for $f(x,v,t)$ the Klimontovitch distribution function, $f(x,v,t) \equiv \sum_{i=1}^N \delta(x-x_i,v-v_i)$, where the sum is over all electrons. However, this would require to solve the $N$-body problem, which is way out of the scope of this paper. Consequently, we use for $f(x,v,t)$ the Vlasovian distribution function, which is not explicitly written in terms of $x_i$ and $v_i$. Nevertheless, when writing Lagrange equations for the EPW electric field, it is important to remember that $f(x,v,t)$ has to be considered as a function of the electrons coordinates. This is particularly important as regards the trapped electrons, whose coordinates can only be defined with respect to the $O$-point of their trapping island. This entails an explicit $\psi$-dependence of the trapped distribution function, which is obvious from Eq.~(\ref{A6b}), and which is further discussed in the Paragraph just above Eq.~(\ref{C3}).

 Now, one may choose to express the Lagrangian for wave-particle interaction in any set of canonically conjugated variables, and, in the near-adiabatic regime, we clearly shift to action-angle variables, $(\theta,I)$. Since we only look here for an expression of $L$ at zero order in the fields variations, we implicitly use the zero-order expressions of $I$ and $\theta$. Hence, in the integrals~Eq.~(\ref{A9})~and~Eq.~(\ref{A9b}), the wave number, amplitude and frequency are assumed to remain constant within one period. For the untrapped electrons, this makes $I$ and $\theta$ local functions of the EPW electric field, while, for the trapped electrons, the non locality is only in $u_0$ and $\eta_0$. Now, from Eq.~(\ref{A4}), in $(\theta,I)$ variables, the Lagrangian for wave-particle interaction is, $\tilde{L}_e=I\dot{\theta}-H_e(I,\theta,t)+\partial g/\partial t\vert_I$. Henceforth, we neglect the term $\partial g/\partial t$, because we look for an averaged variational principle and, as discussed in the Appendix~\ref{A}, over one period, $\partial g/\partial t$ averages out to zero. Then, one may approximate $\dot{\theta}$ by $\Omega \equiv \partial_I \mh$, where $\mh$ is defined by Eq.~(\ref{A7}), so that $\tilde{L}_e \approx I\Omega - H_e$. 

Now, let us denote by $\tilde{f}_u$ the distribution function of the untrapped electrons in action-angle variables, defined such that $\tilde{f}dId \theta=fdxdv$ (then $\tilde{f}=f/m$). As long as $I$ remains nearly conserved, $\tilde{f}_u$ varies very slowly with $\theta$. Actually, in the near-adiabatic regime, the $\theta$-dependence of $\tilde{f}_u$ is only due to the small plasma inhomogeneity. Therefore, $\tilde{f}_u $ reads, $\tilde{f}_u \equiv \tilde{f}_u[x(\theta,I),I,t]$, and varies very slowly with $x$. As for the Hamiltonian, $H_e$, we know from Appendix~\ref{A} that, $H_e(I,\theta,t)=\mH_e[\psi(I,\theta),I,t]$, where $\mH_e$ varies very slowly with $\psi$. Now, in order to use the Lagrangian density, $L$, to derive the field equations, these fields must clearly be expressed as functions of $x$ (and not as functions of $\theta$ and $I$). Hence, we introduce $\mH_t$ such that $\mH_t(x,I,t) \equiv \mH_e[\psi(x,t),I,t]$, and the contribution to $L$ from the untrapped electrons, which we denote by $L_u$, is
\begin{equation}
\label{C2}
L_u(x_0,t) \approx \int [I\Omega-\mH_t(x_0,I,t)] f_u(x_0,I,t) \left.\frac{\partial \theta}{\partial x}\right\vert_{\theta_0}dI,
\end{equation}
where $\theta_0$ is such that $x(\theta_0,I)=x_0$. Now, let us note that $\partial_x \theta=k \partial_\psi\theta$, and let us denote by $l_u$ the integrand in Eq.~(\ref{C2}). Within the multiple-scale approach, $l_u$ reads, $l_u\equiv l_u(\psi,x_0,t)$, and is periodic in $\psi$. More precisely, in $l_u$, the factor $k[I\Omega-\mH_t(x_0,I,t)] f_u(x_0,I,t)$ is to be considered as a function of $x_0$ only. Hence, it remains fixed when, following Whitham theory, one averages over $\psi$ at a fixed value of $x_0$. As for $\partial_\psi \theta$, it is such that $\langle \partial_\psi \theta \rangle=1$, so that any of its derivative averages out to zero. Consequently, $\langle \partial_{xk}ÊL_u -\partial_{t\omega}L_uÊ\rangle=\partial_{xk} \langle L_u \rangle - \partial_{t\omega} \langle L_u \rangle$, and this result is just equivalent to that derived by Whitham~\cite{whitham}. 

As regards the trapped electrons, similarly, one finds $L_eÊ\approx I\Omega-H_e$. However, as outlined in Appendix~\ref{A}, for trapped electrons, one has to introduce one set of action-angle variables per resonance (i.e., per trapping island). Consequently, unlike for untrapped electrons, the distribution function in $(I,\theta)$ cannot be defined continuously in space. For each $x_0$, enters \textit{the} distribution function, $\tilde{f}_t$, corresponding to \textit{the}~$O$-point which is closest to $x_0$, and this $\tilde{f}_t$ is entirely defined by the value, $\psi_O$, of the eikonal at this $O$-point. Hence, about a given location, $x_0$, the trapped distribution function reads, $\tilde{f}_t \equiv \tilde{f}_t(\psi_O,I)$. Similarly, within one wavelength, the Hamiltonian $H_e$ is $\theta$-independent, and is also defined by $\psi_O$. Just like for the untrapped electrons, let us introduce $\mH_t(x_O,I,t)\equiv \mH_e[\psi_O(x_O,t),I,t]$. Then, the contribution, $L_t$, from the trapped electrons to $L$ is,
 \begin{equation}
\label{C3}
L_t(x_0,t) \approx \int [I\Omega-\mH_t(x_O,I,t)] \tilde{f}_t(\psi_O,I,t) \left.\frac{\partial \theta}{\partial x}\right\vert_{\theta_0}dI.
\end{equation}
It is clear from Eq.~(\ref{C3}) that, when calculating the variations of the action, $S \equiv \int L dxdt$, with respect to $\psi$, one has to account for the $\psi$-dependence of $\tilde{f}_t$. $\mH_t$ and $\tilde{f}_t$ are step functions, they are constant within one trapping island. Hence, when calculated between two $O$-points, $O_1$ and $O_2$, separated by the $X$-point, $\mX$, $\partial_x \mH_t=\delta(x-x_{\mX})[\mH_t(x_{O_2})-\mH_t(x_{O_1})] \equiv \delta(x-x_{\mX}) \delta \mH_t$. Similarly, $\partial_\psi \tilde{f}_t=\delta(\psi-\psi_{\mX})[\tilde{f}_t(\psi_{O_2})-\tilde{f}_t(\psi_{O_1})] \equiv \delta(\psi-\psi_{\mX}) \delta \tilde{f}_t$. Considering, now, $\mH_t$ and $\tilde{f}_t$ as continuous space functions, $\mH_t \equiv \mH_t(x)$ and $\tilde{f}_t \equiv \tilde{f}_t(\psi)$, at the order where we work, $\delta \mH_t \approx (2\pi/k) \partial_x \mH_t$ and $\delta \tilde{f}_t \approx (2\pi)^{-1} \partial_\psi \tilde{f}_t$. Therefore, when averaging the variations of $S$ over one wavelength, one may replace in Eq.~(\ref{C3}), $\mH_t(x_O)$ by $\mH_t(x)$ and $\tilde{f}_t(\psi_O)$ by $\tilde{f}_t(\psi)$, but this is an approximation, only valid at first order. Then, within this approximation, and just like for the untrapped electrons, one finds $\langle \partial_{xk}ÊL_t -\partial_{t\omega}L_tÊ\rangle=\partial_{xk} \langle L_t \rangle - \partial_{t\omega} \langle L_t \rangle$. Similarly, $\langle \partial_\psi L_t \rangle=\partial_\psi \langle L_t \rangle$. 

Using the results derived for the trapped and untrapped electrons we conclude that,  $\partial_{xk}Ê\lL-\partial_{t\omega}\lL\approx\partial_\psi \lL$ (since it is clear that $\langle \partial_\psi E^2 \rangle=0$). Unlike in Whitham theory, we do not get an exact equality, the equation is only true at lowest order in the variations of the fields, and, in particular, of $\tilde{f}_t$. 

Let us now discuss Lagrange equations for the amplitudes, $\Phi_n$, of the harmonics of the scalar potential, or for the amplitude, $\delta A$, of the vector potential. Each of these amplitudes varies slowly in space and time. Therefore, within the multiple-scale approach, they are to be considered as $\psi$-independent. Then, using the same derivation as before, one concludes that $\partial_x[\partial \lL/\partial \mF_x]+\partial_t[\partial \lL/\partial \mF_t]=\partial_{\mF}Ê\lL$, where $\mF$ is, either, any of the $\Phi_n$, or is $\delta A$, and where $\mF_w=\partial_w \mF$. 

\section{Lagrange equations for a driven wave}
\label{D}
\setcounter{app}{4}
Let us assume that the plasma wave is driven by the sinusoidal electrostatic field, $E_d \cos(\psi_d)$, and let us write the EPW electric field 
\begin{equation}
E=\sum_{n \geq 1}ÊE_n \sin(\psi_n+\delta \psi_n)+\bar{E}(x,t),
\end{equation}
where $\bar{E}(x,t) \equiv \langle E \rangle$ slowly depends on space and time. If $\delta \psi \equiv\psi_d-\psi$ varies so slowly that it may be considered constant over one wavelength, one needs to include the driving field in the Lagrangian density. Indeed, this field reads as an explicit function of $\psi$. Hence, compared to a freely propagating wave, one just needs to change, $\varepsilon_0 E^2/2$ into $\varepsilon_0 [E+E_d \cos(\psi_d)]^2/2$. Since,
\begin{equation}
\label{D1}
\langle [E+E_d \cos(\psi_d)]^2 \rangle= \langle E^2 \rangle + \langle E_d^2 \rangle +E_1E_d \sin(\psi-\psi_d),
\end{equation}
one finds an extra term in $\partial \mL/\partial \psi$, which is just $\varepsilon_0 E_1E_d \cos(\delta \psi)/2$, and which needs to be substracted to the right-hand side of~\reso. In 3-D, this leads to~\res. 

Moreover, if the EPW is laser driven by SRS, then, it is well known that the driving field derives from the so-called ponderomotive potential (see, for example, Ref.~\cite{benisti07}). Consequently, the results derived in this Appendix applies to the important issue of SRS.

\section{Adiabatic formulas for a sinusoidal wave}
\label{E}
\setcounter{app}{5}
In this Appendix, we give explicit expressions for $\mH_u$, $\mH_t$, and for their derivatives, when the wave is sinusoidal, $E=E_1\sin(\psi)$. Moreover, we use $\Phi_1 \equiv eE_1/k$, $k$ and $\omega$ as independent variables, and we only focus on the derivatives with respect to $\Phi_1$ and to $\omega$, whose expressions may be tested numerically. Furthermore, we restrict to the situation when $\vert dv_\phi/dt \vert >(4/m\pi)d\sqrt{\Phi_1}/dt$, so that the range in action of the trapped and untrapped electrons only depends on the local value of $\Phi_1$. More precisely, let us introduce, for the untrapped electrons,
\begin{equation}
\label{E00}
V_u \equiv P/m-v_\phi,
\end{equation}
and, for the trapped electrons,
\begin{equation}
\label{E01}
V_t \equiv kI_t/m-v_{\phi_0},
\end{equation}
where $v_{\phi_0}$ is the value assumed by the phase velocity when the electron has been trapped. Then, $\vert V_u\vertÊ\geq (4/\pi) \sqrt{\Phi_1/m}$ while $\vert V_t \vert \leq (4/\pi) \sqrt{\Phi_1/m}$, so that the Routhian density introduced in Paragraph~\ref{III.2}, and yielding the fields equations, reads,
\begin{equation}
\label{E02}
\mL = \frac{k^2\varepsilon_0\Phi_1^2}{4e^2}-\int_{\vert V_u \vert \geq \frac{4}{\pi} \sqrt{\frac{\Phi_1}{m}}} F_u(v_\phi+V_u)\mH_udV_u-\int_{-\frac{4}{\pi} \sqrt{\frac{\Phi_1}{m}}}^{\frac{4}{\pi} \sqrt{\frac{\Phi_1}{m}}}F_t(v_\phi+V_t)\mH_tdV_t,
\end{equation}
where the distribution functions, $F_u$ and $F_t$ are such that $F_u(v_\phi+V_u)dV_u=f_u(P)dP$ and $F_t(v_\phi+V_t)dV_t=kf_t(I_t)dI_t$. 

As usual regarding sinusoidal waves, we introduce for untrapped electrons, $\zeta_u \equiv (\mh+\Phi_1)/2\mh$, so that 
\begin{eqnarray}
\label{E1}
\nonumber
\mH_u&=& \frac{2-\zeta_u}{\zeta_u}\Phi_1+Pv_\phi-mv_\phi^2-eAv_\phi \\
&=& \frac{2-\zeta_u}{\zeta_u}\Phi_1-\frac{mV_u^2}{2}-\frac{P^2}{2m}-eAv_\phi,
\end{eqnarray}
while, for trapped electrons, we introduce $\zeta_t \equiv 2\mh/(\mh+\Phi_1)$, which leads to 
\begin{equation}
\label{E2}
\mH_t=(2\zeta_t-1)\Phi_1-m(v_\phi-eA/m)^2/2-e^2A^2/2m.
\end{equation}
Moreover, as shown in Ref.~\cite{benisti07}, 
\begin{eqnarray}
\label{E3}
V_u &=& \eta_u\frac{4}{\pi}\sqrt{\frac{\Phi_1}{m}} \frac{K_2(\zeta_u)}{\sqrt{\zeta_u}}, \\
\label{E4}
V_t &=& \eta_t\frac{4}{\pi}\sqrt{\frac{\Phi_1}{m}} \left[K_2(\zeta_t)+(\zeta_t-1)K_1(\zeta_t)Ê\right],
\end{eqnarray}
where $\eta_u$ and $\eta_t$ are, respectively, the sign of $V_u$ and $V_t$, and where $K_1$ and $K_2$ are, respectively, the Jacobian elliptic integral of first and second kind~\cite{abramowitz}.

From Eq.~(\ref{E3}) it is easily found that $\partial_{\Phi_1}Ê\zeta_u= \zeta_u K_2/\Phi_1K_1$, which yields,
\begin{equation}
\label{E5}
\frac{\partial \mH_u}{\partial \Phi_1}=-\left[1+\frac{2}{\zeta_u}\left(\frac{K_2}{K_1}-1 \right)Ê\right].
\end{equation}
Similarly, from Eq.~(\ref{E4}), $\partial_{\Phi_1}Ê\zeta_t=-\left[(\zeta_t-1)K_1+K_2Ê\right]/K_1\Phi_1$, which yields,
\begin{equation}
\label{E6}
\frac{\partial \mH_t}{\partial \Phi_1}=-\left[\frac{2K_2}{K_1}-1\right].
\end{equation}
Therefore, the Lagrange equation $\partial_{\Phi_1}Ê\mL=0$ reads,
\begin{eqnarray}
\nonumber
1+\frac{2e^2}{\varepsilon_0 \Phi_1 k^2}&&\left\{\int_{\vert V_u \vert \geq \frac{4}{\pi}\sqrt{\frac{\Phi_1}{m}}} \left[1+\frac{2}{\zeta_u}\left(\frac{K_2}{K_1}-1 \right)Ê\right]F_u(v_\phi+V_u)dV_uÊ\right.\\
\label{E7}
&&\left.+\int_{-\frac{4}{\pi}\sqrt{\frac{\Phi_1}{m}}}^{\frac{4}{\pi}\sqrt{\frac{\Phi_1}{m}}}\left[\frac{2K_2}{K_1}-1\right]F_t(v_\phi+V_t)dV_t \right\}=0,
\end{eqnarray}
where $\zeta_u$ and $\zeta_t$ are, respectively, considered as functions of $V_u$ and $V_t$ according to Eqs.~(\ref{E3})~and~(\ref{E4}).

Note that, when calculating $\partial_{\Phi_1}Ê\mL$, one must not account for the derivatives of the integral boundaries in Eq.~(\ref{E02}). This prescription, which holds whatever the derivative, directly follows from the derivation of Lagrange equations given in Paragraph~\ref{III.2.2}.  Physically, the abrupt distinction between the responses from trapped and untrapped electrons, coming from the adiabatic model, is spurious. Clearly, the electron response should be a continuous function of the action, the expression obtained from the untrapped electrons being smoothly connected to that derived from the trapped ones within a range of action of the order of $m\vert \gamma \vert/k^2$, where $\gamma$ is the rate of variation of the wave amplitude.  Indeed, as discussed in Paragraph~\ref{new3}, trapping is an ambiguous notion for a wave whose amplitude varies, and, within a range in action of the order of $m\vert \gamma\vert /k^2$ about $(4m/\pi)\sqrt{m\Phi_1}$, the electron orbit may neither be considered as trapped nor untrapped.

Now, using the identities,
\begin{eqnarray}
\label{E8}
\int_{\vert V_u \vert \geq \frac{4}{\pi}\sqrt{\frac{\Phi_1}{m}}} \left[1+\frac{2}{\zeta_u}\left(\frac{K_2}{K_1}-1 \right)Ê\right]dV_uÊ &=&\frac{8}{3\pi}Ê\sqrt{\frac{\Phi_1}{m}}\\
\label{E9}
\int_{-\frac{4}{\pi}\sqrt{\frac{\Phi_1}{m}}}^{\frac{4}{\pi}\sqrt{\frac{\Phi_1}{m}}}\left[\frac{2K_2}{K_1}-1\right]dV_t&=&-\frac{8}{3\pi}Ê\sqrt{\frac{\Phi_1}{m}},
\end{eqnarray}
Eq.~(\ref{E7}) reads,
\begin{eqnarray}
\nonumber
1+\frac{2e^2}{\varepsilon_0 \Phi_1 k^2}&&\left\{\int_{\frac{4}{\pi}\sqrt{\frac{\Phi_1}{m}}}^{+\infty} \left[1+\frac{2}{\zeta_u}\left(\frac{K_2}{K_1}-1 \right)Ê\right]\delta_+F_u(V_u)dV_uÊ\right.\\
\nonumber
&&\left.+\int_{0}^{\frac{4}{\pi}\sqrt{\frac{\Phi_1}{m}}}\left[\frac{2K_2}{K_1}-1\right]\delta_+F_t(V_t)dV_t+\frac{8}{3\pi}Ê\left[F_u(v_\phi)-F_t(v_\phi)Ê\right] Ê\sqrt{\frac{\Phi_1}{m}}\right\}=0,\\
\label{E10}
\end{eqnarray}
where $\delta_+F \equiv F(v_\phi+V)+F(v_\phi-V)-2F(v_\phi)$. Eq.~(\ref{E10}) is the nonlinear dispersion relation of a nearly sinusoidal wave in the near-adiabatic regime. \\

Let us now evaluate $\partial^2_{\Phi_1\omega} \mL$, which yields the term proportional to $\Phi_1\partial_t \Phi_1$ in the envelope equation. Using, $\partial_\omega \mH_u=k^{-1}Ê\partial_{v_\phi}\mH_u$, and, from Eq.~(\ref{E00}), $\partial_{v_\phi} \mH_u=-\partial_{V_u} \mH_u$, one straightforwardly finds,
\begin{equation}
\label{E11} 
-\frac{\partial^2 \mL}{\partial \Phi_1 \partial \omega}=-\frac{1}{k} \int^{+\infty}_{\frac{4}{\pi}\sqrt{\frac{\Phi_1}{m}}} \left[F_u(v_\phi+V_u)+F_u(v_\phi-V_u)Ê\right]\frac{\partial}{\partial V_u} \left[1+\frac{2}{\zeta_u}\left(\frac{K_2}{K_1}-1 \right)Ê\right]dV_u.
\end{equation}
Moreover, taking advantage of the indentity,
\begin{equation}
\label{E12}
\int^{+\infty}_{\frac{4}{\pi}\sqrt{\frac{\Phi_1}{m}}} V_u \frac{\partial}{\partial V_u} \left[1+\frac{2}{\zeta_u}\left(\frac{K_2}{K_1}-1 \right)Ê\right]dV_u=\frac{16}{3\pi}\sqrt{\frac{\Phi_1}{m}},
\end{equation}
Eq.~(\ref{E11}) reads,
\begin{equation}
\label{E13}
-\frac{\partial^2 \mL}{\partial \Phi_1\partial \omega}=-\frac{1}{k} \int^{+\infty}_{\frac{4}{\pi}\sqrt{\frac{\Phi_1}{m}}} \delta_-F_u(V_u)\frac{\partial}{\partial V_u} \left[1+\frac{2}{\zeta_u}\left(\frac{K_2}{K_1}-1 \right)Ê\right]dV_u-\frac{32}{3\pi k}\sqrt{\frac{\Phi_1}{m}}F'_u(v_\phi),
\end{equation}
where $\delta_-F(V) \equiv F(v_\phi+V)-F(v_\phi-V)-2VF'(V)$, and $F'(V) \equiv \partial_{V}F$. \\

For the purpose of discussing how the wave action should best be defined, we now calculate $\partial_{\Phi_1\omega_{all}} \mL$, which we define the following way. For the term involving the untrapped electrons, we just define $\partial_{\Phi_1\omega_{all}} \mH_u \equiv \partial_{\Phi_1\omega} \mH_u$. As regards the trapped electrons, we introduced for each trapped orbit, $v_{\phi_0} \equiv \omega_0/k_0$, the value assumed by the wave phase velocity when the orbit was being trapped. Then, we define $\partial_{\Phi_1\omega_{all}} \mH_t \equiv \partial_{\Phi_1\omega_0} \mH_t$. This yields,
\begin{eqnarray}
\nonumber
-\frac{\partial^2 \mL}{\partial \Phi_1 \partial \omega_{all}}=&&-\frac{1}{k} \int^{+\infty}_{\frac{4}{\pi}\sqrt{\frac{\Phi_1}{m}}} \left[F_u(v_\phi+V_u)+F_u(v_\phi-V_u)Ê\right]\frac{\partial}{\partial V_u} \left[1+\frac{2}{\zeta_u}\left(\frac{K_2}{K_1}-1 \right)Ê\right]dV_u \\
\label{E13b}
&&-\frac{1}{k} \int^{\frac{4}{\pi}\sqrt{\frac{\Phi_1}{m}}}_{-\frac{4}{\pi}\sqrt{\frac{\Phi_1}{m}}}\left[F_t(v_\phi+V_t)+F_t(v_\phi-V_t)Ê\right]\frac{\partial}{\partial V_t} \left[ \frac{2K_2}{K_1}-1Ê\right]dV_t.
\end{eqnarray}
Using Eq.~(\ref{E12}) and the identity,
\begin{equation}
\label{E13c}
\int^{\frac{4}{\pi}\sqrt{\frac{\Phi_1}{m}}}_{-\frac{4}{\pi}\sqrt{\frac{\Phi_1}{m}}} V_t \frac{\partial}{\partial V_t}  \left[ \frac{2K_2}{K_1}-1Ê\right]dV_t=-\frac{16}{3\pi}\sqrt{\frac{\Phi_1}{m}},
\end{equation}
Eq.~(\ref{E13b}) reads,
\begin{eqnarray}
\nonumber
-\frac{\partial^2 \mL}{\partial \Phi_1 \partial \omega_{all}}=&&-\int^{+\infty}_{\frac{4}{\pi}\sqrt{\frac{\Phi_1}{m}}} \frac{\delta_-F_u(V_u)}{k}\frac{\partial}{\partial V_u} \left[1+\frac{2}{\zeta_u}\left(\frac{K_2}{K_1}-1 \right)Ê\right]dV_u \\
\nonumber
&&-\int^{\frac{4}{\pi}\sqrt{\frac{\Phi_1}{m}}}_{-\frac{4}{\pi}\sqrt{\frac{\Phi_1}{m}}} \frac{\delta_-F_t(V_t)}{k}\frac{\partial}{\partial V_t} \left[ \frac{2K_2}{K_1}-1Ê\right]dV_t \\
\label{E13e}
&&-\frac{32}{3\pi}\sqrt{\frac{\Phi_1}{m}}\left[F'_u(v_\phi)-F'_t(v_\phi)Ê\right].
\end{eqnarray}
\newline

Let us now come to $\partial_{\omega^2}Ê\mL$, which yields the term proportional to $\Phi_1^2\partial_t \omega$ in the envelope equation. From Eq.~(\ref{E02}), 
\begin{equation}
\label{E14}
-\frac{\partial^2\mL}{\partial \omega^2}=-\int^{+\infty}_{\frac{4}{\pi}\sqrt{\frac{\Phi_1}{m}}} \frac{F_u(v_\phi+V_u)+F_u(v_\phi-V_u)}{k^2}\frac{\partial^2\mH_u}{\partial V_u^2}dV_u+m\int_{-\frac{4}{\pi}\sqrt{\frac{\Phi_1}{m}}}^{\frac{4}{\pi}\sqrt{\frac{\Phi_1}{m}}}\frac{F_t(v_\phi+V_t)}{k^2}dV_t. 
\end{equation}
Now, it is easily found that
\begin{eqnarray}
\label{E15}
\int^{+\infty}_{\frac{4}{\pi}\sqrt{\frac{\Phi_1}{m}}}\frac{\partial^2\mH_u}{\partial V_u^2}dV_u&=&-\frac{4m}{\pi}\sqrt{\frac{\Phi_1}{m}} \\
\int^{+\infty}_{\frac{4}{\pi}\sqrt{\frac{\Phi_1}{m}}}\frac{V_u^2}{2}\frac{\partial^2\mH_u}{\partial V_u^2}dV_u&=&-\frac{m}{6}\left(\frac{4\Phi_1}{\pi m} \right)^{3/2}-\frac{32m}{3\pi^2}\left(\frac{\Phi_1}{m} \right)^{3/2},
\end{eqnarray}
so that,
\begin{eqnarray}
\nonumber
-\frac{\partial^2\mL}{\partial \omega^2}=&&-\int^{+\infty}_{\frac{4}{\pi}\sqrt{\frac{\Phi_1}{m}}} \frac{\delta^2_+F_u(V_u)}{k^2}\frac{\partial^2\mH_u}{\partial V_u^2}dV_u\\
\nonumber
&&+m\int_{-\frac{4}{\pi}\sqrt{\frac{\Phi_1}{m}}}^{\frac{4}{\pi}\sqrt{\frac{\Phi_1}{m}}}\frac{F_t(v_\phi+V)-F_u(v_\phi)-(V^2/2)F_u''(v_\phi)}{k^2}dV \\
\label{E16}
&&-\frac{64m}{3\pi^2k^2}\left(\frac{\Phi_1}{m} \right)^{3/2}F''_u(v_\phi),
\end{eqnarray}
where $\delta^2_+F(V) \equiv F(v_\phi+V)+F(v_\phi-V)-2F(v_\phi)-V^2F''(v_\phi)$.


\begin{thebibliography}{10}
\bibitem{comment}
In Ref.~\cite{mouhot}, Mouhot and Villani derived Landau damping from a direct resolution of the Vlasov-Poisson system without making use of the linear approximation. However, their derivation is only valid when the potential amplitude is small enough so that, in practice, the electrons motion is very close to linear.
\bibitem{comment2}
Clearly, for a given field, $\me$, there is more than one choice for the wave amplitude and eikonal. However, within the geometrical optics limit, they are all equivalent and would lead to the same  envelope equation and dispersion relation. Indeed, let us start with the linear regime when the wave is sinusoidal, $\me \equiv \mE \sin(\psi)$. If one makes another choice, $\me \equiv \mE' \sin(\psi')$, then, the zeros of $\sin(\psi)$ and $\sin(\psi')$ are necessarily the same. This entails that  $\psi \approx \psi'$, since the time and space derivatives of $\psi$ and $\psi'$ are slowly-varying. Consequently $\mE \approx \mE'$. Then, once $\psi$ is defined in the linear regime, it is continued in the nonlinear one using the dispersion relation, as described, for example, in Ref.~\cite{lindberg}.
\bibitem{comment4}
Here we assume that electrons experience only a few separatrix crossings. Indeed, we consider the situation where the EPW is located within a finite space domain that electrons cross due to their transverse motion. Hence, chaos induced by numerous separatrix  crossings is excluded here.
\bibitem{note}
The linear susceptibility and the Landau damping rate are indeed recovered provided that $F_u$ has remained very close to the unperturbed distribution function. This approximation is valid in the small amplitude limit since, at lowest order, the term that makes $F_u$ vary in time scales as $\Phi_1^2$.
\end{thebibliography}
\end{document}